\documentclass[a4paper, amsfonts, amssymb, amsmath, reprint, showkeys, nofootinbib, twoside, onecolumn,notitlepage]{revtex4-2}
\def\prd{{Phys.~Rev.~D}}        % Physical Review D
\usepackage{amsmath,amstext}
\usepackage[T1]{fontenc}
\usepackage{amssymb}
\input{epsf}
\usepackage{graphicx}
\usepackage{ae,aecompl}
\usepackage{combelow}

\usepackage{standalone}
\usepackage{hyperref}
\usepackage{amsmath}
\usepackage{amssymb}
\usepackage{mathtools}
\usepackage{bm}
\usepackage{cleveref}
\usepackage{tensor}
\usepackage{braket}
\usepackage{enumitem}
\usepackage{mhchem}
\usepackage{cancel}
\usepackage{tikz}
\usepackage{pgfplots}
\usetikzlibrary{external}
\usepackage{mathrsfs}
\usepackage{ulem}

\usepackage{color}

\newcommand{\spc}{\quad \quad \quad}

\renewcommand{\d}{\mathrm{d}}
\newcommand{\A}{\mathbb{A}}
\renewcommand{\k}{\mathbf{k}}

\def\be{\begin{equation}}
\def\ee{\end{equation}}
\def\beq{\begin{eqnarray}}
\def\eeq{\end{eqnarray}}

\begin{document}
\title{The regime of applicability of Israel-Stewart hydrodynamics}
\author{D.~Wagner$^{1,2}$ and L.~Gavassino$^3$}
\affiliation{$^{1}$Institute for Theoretical Physics, Goethe University,
Max-von-Laue-Str.\ 1, D-60438 Frankfurt am Main, Germany}
\affiliation{$^{2}$Department of Physics, West University of Timi\cb{s}oara, \\
Bd.~Vasile P\^arvan 4, Timi\cb{s}oara 300223, Romania}
\affiliation{
$^3$Department of Mathematics, Vanderbilt University, Nashville, TN, USA
}

\begin{abstract}
Using analytical tools from linear response theory, we systematically assess the accuracy of several microscopic derivations of Israel-Stewart hydrodynamics near local equilibrium. This allows us to ``rank'' the different approaches in decreasing order of accuracy as follows: Inverse Reynolds Dominance (IReD), Denicol-Niemi-Molnár-Rischke (DNMR), second-order gradient expansion, and 14-moment approximation. We find that IReD theory is far superior to Navier-Stokes, being very accurate both in the asymptotic regime (i.e., for slow processes) and in the transient regime (i.e., on timescales comparable to the relaxation time). Also, the high accuracy of DNMR is confirmed, but neglecting second-order terms in the Knudsen number, which would render the equations parabolic, introduces serious systematic errors. Finally, {in most cases}, the second-order gradient expansion (a.k.a. non-resummed BRSSS) is {found} to be more inaccurate than Navier-Stokes in the transient regime. Overall, this analysis shows that Israel-Stewart hydrodynamics is falsifiable, and the relaxation time is observable, {shedding new light on} the debate on the viability of transient hydrodynamics as a well-defined physical theory distinguished from Navier-Stokes.
\end{abstract}

\maketitle

\section{Introduction}

\begin{figure}
\centering
\includegraphics[width=.49\textwidth]{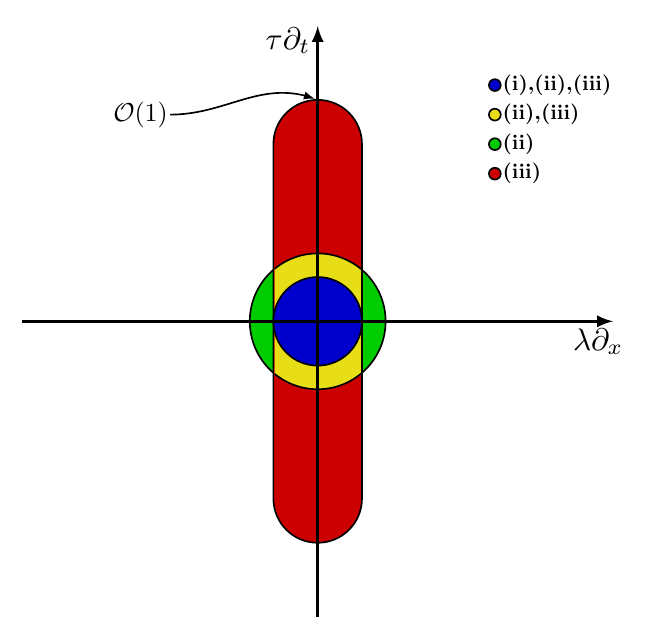}
\caption{{Fix a spacetime event $\mathscr{P}$. Working in the local rest frame of the fluid, call ``$\partial_x$'' the maximum (in magnitude) rate of change of the hydrodynamic fields along all spacelike directions, and call ``$\partial_t$'' the maximum rate of change along the time direction (near $\mathscr{P}$). Compare them with the particles' mean free path $\lambda$ and mean collision time $\tau$. The \textit{regime of applicability} of a hydrodynamic theory (in our case, Israel-Stewart) is the region of the $\{\lambda \partial_x,\tau \partial_t\}$ plane where the predictions of hydrodynamics are assumed accurate. Both interpretations (i) and (ii) treat spacelike derivatives and timelike derivatives on the same footing, and (ii) assumes a slightly larger regime than (i). Interpretation (iii) claims a regime of applicability that extends up to $|\tau \partial_t| \lesssim 1$.}}
\label{fig:applicabluz}
\end{figure}

There is a long-lasting debate over the physical content of Israel-Stewart-type theories \cite{Israel_Stewart_1979,Hishcock1983,OlsonLifsh1990,Baier2008,Denicol2012Boltzmann,GavassinoGENERIC2022} for relativistic viscous hydrodynamics. The three most widespread positions on the matter are summarised below ({see figure \ref{fig:applicabluz}}): 
\begin{itemize}
\item[(i)] Israel-Stewart theories are a ``mathematical trick'' to make relativistic Navier-Stokes hyperbolic, causal, and stable. Their physical content is exactly the same as Navier-Stokes, and all the additional transport coefficients (e.g. the relaxation times) are unobservable UV regulators, which fall outside the regime of applicability of hydrodynamics \cite{Geroch1995,LindblomRelaxation1996,Geroch2001,Kost2000}.
\item[(ii)] Israel-Stewart theories are a ``refinement'' of Navier-Stokes, able to capture the dynamics up to second-order in gradients, from which the term ``second-order hydrodynamics'' comes \cite{Baier2008,Romatschke2017,FlorkowskiReview2018}. Their regime of applicability is essentially the same as that of Navier-Stokes (i.e., small gradients), but the dynamics is captured more accurately.
\item[(iii)] Israel-Stewart theories are an extension of hydrodynamics which captures not only the Navier-Stokes behaviour, but also the dynamics of the slowest non-equilibrium degrees of freedom (namely the first non-hydrodynamic modes \cite{Denicol_Relaxation_2011,Grozdanov2013,GavassinoNonHydro2022,GavassinoFronntiers2021}). Their regime of applicability extends beyond Navier-Stokes, and it can capture also the initial transient dynamics of a fluid, when the slowest non-equilibrium degrees of freedom have not yet equilibrated. From this the term ``transient hydrodynamics'' originates \cite{Israel_Stewart_1979,Denicol2012Boltzmann}. 
\end{itemize}
The tenured practitioner may regard such debate as a merely academic controversy. However, this discussion has a direct impact on theoretical models and numerical simulations, because it affects how we define and compute the Israel-Stewart transport coefficients. If we adopt interpretation (i), then it does not really matter what the actual value of, e.g., the relaxation time is, as long as causality and stability are enforced. If we adopt interpretation (ii), then there is a Kubo formula for all Israel-Stewart coefficients, including the relaxation time. Finally, if we adopt interpretation (iii), then the relaxation time is determined by the first non-hydrodynamic pole of the retarded correlator, and there is no Kubo formula for it \cite{Denicol_Relaxation_2011}.

Let us consider a concrete example of this dilemma: the problem of deriving relativistic hydrodynamics from kinetic theory. Here, one starts from a theory with infinite local degrees freedom, namely all the moments of the phase-space distribution function $f(x^\mu,p^\mu)$, and the goal is to make some approximation to derive a closed system of equations for the Israel-Stewart hydrodynamic fields, which constitute only 14 algebraic degrees of freedom. The best-known approach to this problem is the DNMR procedure \cite{Denicol2012Boltzmann} which, in its original intent, aims to fulfill both interpretations (ii) and (iii). In fact, the authors of \cite{Denicol2012Boltzmann} carry out a rigorous expansion up to second order in gradients, consistent with point (ii). At the same time, they develop an ingenious technique to match the relaxation times with those of the slowest non-hydrodynamic modes, thereby ensuring consistency also with point (iii). Unfortunately, the resulting theory fails to be hyperbolic, opening the doors to causality violations and, therefore, to instabilities \cite{GavassinoSuperluminal2021,GavassinoBounds2023}. For this reason, when the DNMR theory is actually used in numerical simulations, one is forced to artificially remove all the terms that are explicitly of second order in gradients (the terms $\mathcal{K}$, $\mathcal{K}^\mu$, $\mathcal{K}^{\mu \nu}$ \cite{Denicol2012Boltzmann, Molnar:2013lta,Denicol:2012vq}), breaking consistency with point (ii). To fix this problem, an alternative procedure called Inverse Reynolds Dominance (IReD) has been recently proposed \cite{WagnerIReD2022}\footnote{We remark that the method has also been called ``order-of-magnitude approximation'' \cite{Fotakis:2022usk, Rocha:2023hts} and has its origins in nonrelativistic kinetic theory \cite{Struchtrup}.}. The main idea of \cite{WagnerIReD2022} is to ``reabsorb'' the $\mathcal{K}$ terms through a redefinition of the hydrodynamic fields. In this way, nothing is neglected, and agreement with (ii) is restored. The price of this transformation is that the relaxation times are redefined, and they may no longer match the transient dynamics of the gas, thereby breaking (iii). Hence, the choice between IReD and DNMR looks equivalent to the choice between interpretations (ii) and (iii).

A rather open-minded approach to the dilemma may be to suggest that there is no universally preferable interpretation of Israel-Stewart theories, and that the choice between positions (i), (ii), and (iii)  ultimately depends on the details of the system under consideration, and on which observables one is interested in. While such a compromise might seem reasonable at first, it still leaves some issues unresolved. For example, the core position of \citet{Geroch2001}, who advocates for (i), is that as soon as deviations from Navier-Stokes become measurable, all the degrees of freedom of kinetic theory ``appear together'', and hydrodynamics as a  whole breaks down. Hence, any ``Israel-Stewart effect'' is fundamentally undetectable, because it is impossible to disentangle it from all the microscopic deviations from hydrodynamics. This (rather extreme) position is at odds with a large part of the recent literature \cite{Denicol:2014xca,Strickland:2014pga,Denicol:2014tha,Alqahtani:2017mhy,Jaiswal:2019cju,Dash:2022xkz,GavassinoFarFromBulk2023,GavassinoBurgers2023}. For this reason, it is now necessary to put Geroch's claims to the test. Furthermore, even assuming that position (i) is not always valid, still we need a clear-cut criterion to decide between interpretations (ii) and (iii) for a given system. 

This article aims to {make some further steps towards settling} all such matters. We will use a solvable linear-response model to discuss the accuracy of the most widespread formulations of Israel-Stewart hydrodynamics in different dynamical regimes. This will allow us to assess the viability of interpretations (i), (ii), and (iii) in realistic situations. Our analysis is restricted to fluid flows that are close to local equilibrium, because a proper formulation of far-from-equilibrium hydrodynamics does not exist, except in some very specific situations \cite{Strickland:2014pga,GavassinoFarFromBulk2023}.

Throughout the article, we adopt the metric signature $({+}{-}{-}{-})$, and work in natural units: $c=k_B=\hbar=1$. We adopt Einstein's convention for spacetime indices ($\mu,\nu,\ldots$).

\section{Mathematical formulation of the problem}

\subsection{A simple linear-response model}\label{lineisaline}

\begin{figure}
\centering
\includegraphics[width=.38\textwidth]{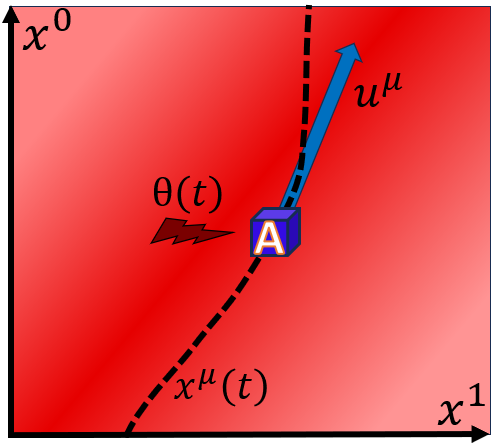}
\caption{{Schematic representation of our linear-response framework.  In a full ``3+1 dimensional'' description, the non-equilibrium degrees of freedom $\mathbb{A}(x^\mu)$ obey some linearised partial differential equations of the form $\tau  u^\mu \partial_\mu  \mathbb{A}+\mathbb{A}=\text{``spatial gradients''}$. However, we can effectively reduce the problem to ``0+1 dimensional'' by tracking the evolution of $\mathbb{A}$ along a single worldline $x^\mu(t)$ (dashed line), which is an integral curve of the velocity field $u^\mu$ (blue arrow). In this dimensional reduction, we lose information. In fact, by focusing on a single worldline, we are forced to treat the spatial gradients as externally given, converting them into an external force $\theta(t)$. This leads to a model where the fluid element (blue box) travels in a dynamically active environment.}}
\label{fig:MDL}
\end{figure}

Let us formulate the problem in an abstract way. We can picture every element of fluid as a tiny thermodynamic system, having a large list of internal non-equilibrium degrees of freedom, $\mathbb{A}=\{\mathbb{A}^1,\mathbb{A}^2,\ldots \}^\mathrm{T}$ (which we will in the following also call ``affinities'' \cite{peliti_book}), undergoing some complicated coupled dynamics. In hydrodynamics, we are not interested in explicitly tracking all these internal variables. Instead, we only want to know the evolution of a single viscous flux $\Pi(\mathbb{A})$, which describes the collective influence of $\mathbb{A}$ on the transport of some conserved quantity. For example, we may take $\Pi$ to be the bulk stress, which affects the transport of linear momentum. If we follow the evolution of a single element of fluid, we can express all our variables as functions of the proper time $t$ along the element's worldline: $\mathbb{A}(t)$, $\Pi(t)$. Now, the time derivative of $\mathbb{A}(t)$, which we call $\dot{\mathbb{A}}(t)$, depends on the state of the fluid element, namely on $\mathbb{A}(t)$ itself, but also on the interaction of the fluid element with the surroundings, since the fluid element is an open system {(see figure \ref{fig:MDL})}. For clarity, we will assume that the action of the environment on $\mathbb{A}(t)$ can be parameterized through a single driving force $\theta(t)$ related to some spatial gradient (in the example of bulk viscosity, we may identify $\theta$ with the expansion rate $\nabla_\mu u^\mu$). Then, linearising the dynamics around local equilibrium, which we take to be the state $\mathbb{A}=0$, the most general model we can construct is\footnote{{In Appendix \ref{app:kin_theory}, we clarify the connection of our simple model to the system of moment equations in kinetic theory.}}
\begin{equation}\label{SuperModel}
\begin{split}
& \tau \dot{\mathbb{A}}+\mathbb{A}=\kappa \theta \, , \\
& \Pi= -\gamma \mathbb{A} \, . \\
\end{split}
\end{equation}
Here, $\tau$ is a square matrix, called ``relaxation matrix'', whose eigenvalues are positive by thermodynamic stability\footnote{Strictly speaking, thermodynamic stability only requires the real part of the eigenvalues to be positive. Hence, depending on the parity of the degrees of freedom under time reversal, one is in principle allowed to have ``complex relaxation times'' \cite{GavassinoNonHydro2022}. Indeed, this is expected to happen in most holographic strongly coupled plasmas \cite{Heller2014}. However, for such systems, the Israel-Stewart theory is not applicable \cite{Denicol_Relaxation_2011}. Hence, focusing on situations that are closer to kinetic theory, we will assume that the eigenvalues of $\tau$ are real and strictly positive.}. The column vector $\kappa=\{\kappa^1,\kappa^2,\ldots\}^\mathrm{T}$ is the dynamical susceptibility of $\mathbb{A}$ to the local gradient $\theta$. In the second equation, we have Taylor-expanded the function $\Pi(\mathbb{A})$ to linear order in $\mathbb{A}$, so that the row vector $\gamma=\{\gamma_1,\gamma_2,\ldots\}$ is just $-\partial_{\mathbb{A}}\Pi(\mathbb{A}{=}0)$. Furthermore, we used that $\Pi(\mathbb{A}{=}0)=0$. In principle, $\tau$, $\kappa$, and $\gamma$ may be functions of the proper time, because they depend on the thermodynamic conditions of the fluid element (e.g. on its energy). Here, we will treat them as constants, for simplicity.\}

Model \eqref{SuperModel} can be solved analytically, and it gives 
\begin{equation}\label{SuperPrediction}
\begin{split}
\mathbb{A}(t)={}&\int_{-\infty}^t e^{-\tau^{-1}(t-t')} \tau^{-1}\kappa \theta(t')\d t' \, , \\
\Pi(t)={}&{-} \! \! \int_{-\infty}^t \gamma e^{-\tau^{-1}(t-t')} \tau^{-1}\kappa \theta(t')\d t' \, . \\
\end{split}
\end{equation}
We are interested in the second equation, which tells us that the fluid elements, when subject to a time-varying driving force $\theta(t)$, will possess an induced viscous flux $\Pi(t)$, which can be expressed in the form $\Pi(t)={-}\int_{\mathbb{R}} G(t-t')\theta(t')\d t'$, with linear-response (retarded) Green function
\begin{equation}\label{gigia}
    G(t) =\Theta(t) \, \gamma e^{-\tau^{-1}t} \tau^{-1}\kappa \, , 
\end{equation}
where $\Theta(t)$ is the Heaviside step function. Assuming, as is usually the case, that $\tau$ is diagonalizable, i.e., $\tau=\sum_n \tau_n \mathcal{P}_n$, with eigenvalues $\tau_n$ and eigenprojectors $\mathcal{P}_n$ (recall: $\sum_n \mathcal{P}_n{=}\mathbb{I}$, $\mathcal{P}_m \mathcal{P}_n{=}\delta_{mn}\mathcal{P}_n$), we can express $G(t)$ in the form
\begin{equation}\label{Green!}
G(t)= \Theta(t) \sum_n \dfrac{\zeta_n}{\tau_n} e^{-t/\tau_n} \, , \spc \text{with }\, \zeta_n = \gamma \mathcal{P}_n \kappa \, .    
\end{equation}
Recalling that $\tau_n{>}0$, we see that $\Pi(t)$ is a sum of independent relaxing contributions $\Pi_n(t) {\propto} \, e^{-t/\tau_n}$. The coefficient $\zeta_n$ quantifies the susceptibility of $\Pi_n$ to the driving force $\theta$. The larger $\zeta_n$, the more $\Pi_n$ is excited by the local gradients.

In the case of kinetic theory, $\{\A^1,\A^2,\ldots \}$ may be interpreted as the infinite tower of moments of the phase-space distribution function $f(x^\mu,p^\mu)$, and $\theta$ is some gradient of the flow velocity. Then, $\Pi_n$ is the contribution to the viscous stress associated with the $n$-th eigenmode of the collision kernel.

\subsection{The Israel-Stewart approximation}

The Israel-Stewart theory is a drastic approximation of model \eqref{SuperModel}. It replaces the system of coupled differential equations for the degrees of freedom $\mathbb{A}$ with a single relaxation equation for $\Pi$ itself:
\begin{equation}\label{LowerModel}
    \tau_\Pi \dot{\Pi} + \Pi = -\zeta \theta \, .
\end{equation}
The underlying assumption is that all the complicated dynamics of $\{\mathbb{A}^1,\mathbb{A}^2,\ldots\}$ can be ``traced out'', and reabsorbed into two transport coefficients, namely $\tau_\Pi$ and $\zeta$. Now, it is evident that the Israel-Stewart theory cannot agree with \eqref{SuperModel} in an exact manner, because the linear-response Green function of \eqref{LowerModel} is
\begin{equation}\label{Ghisa}
   G_{\mathrm{IS}}(t)= \Theta(t) \dfrac{\zeta}{\tau_\Pi} e^{-t/\tau_{\Pi}} \, , 
\end{equation}
which differs from \eqref{Green!} whenever there are at least two relaxation times. However, by appropriately tuning the values of $\tau_\Pi$ and $\zeta$, we may replicate some essential features of \eqref{Green!}. This is where the debate outlined in the introduction comes about. In fact, depending on which interpretation one chooses, there may be different features of $G$ one may want $G_{\mathrm{IS}}$ to replicate, and this leads to different prescriptions for the values of $\tau_\Pi$ and $\zeta$. For example, if one adopts interpretation (i), then only the Navier-Stokes limit $\Pi \approx -\zeta \theta$ matters, meaning that $\tau_\Pi$ can be chosen freely. If one instead follows the DNMR approach, then $\tau_\Pi$ should replicate the evolution of the slowest non-equilibrium degree of freedom, and we should therefore set $\tau_\Pi = \max \{ \tau_n \}$.

\subsection{Our solution}

Let us anticipate the central result of this article: \textit{There is a prescription for $\tau_\Pi$ and $\zeta$ that can accommodate for both interpretations (ii) and (iii) simultaneously, and this prescription is not DNMR in its original formulation \cite{Denicol2012Boltzmann}, but the more recent IReD formulation \cite{WagnerIReD2022}. Within such a framework, the Israel-Stewart predictions are significantly more accurate than the Navier-Stokes ones, thereby falsifying Geroch's claim that Israel-Stewart effects are always undetectable \cite{Geroch2001}\footnote{We would like to point out that with this we are not claiming that interpretation (i) is always incorrect and should be discarded. On the contrary, interpretation (i) is the most convenient and pragmatic approach in all those systems where Navier-Stokes (and therefore BDNK \cite{Bemfica2017TheFirst,Kovtun2019,BemficaDNDefinitivo2020}) is assumed applicable. However, it is incorrect to claim that Israel-Stewart is always only as accurate as Navier-Stokes.}.} Below, we provide some quick analytic arguments in support of the above statements, which are then corroborated by numerical studies in the upcoming sections.

According to interpretation (ii), the ``correct'' Israel-Stewart theory should agree to second order in spacetime gradients with the exact model. Considering that $\theta$ is already of first order in gradients, we only need to make sure that, if we Fourier-transform to frequency space, the Israel-Stewart relationship between $\Pi(\omega)$ and $\theta(\omega)$ agrees with the exact relationship derived from model \eqref{SuperModel}, within an error of order $\omega^2$. Hence, we must match the coefficients of the $\omega-$expansions below:
\begin{equation}\label{bornone}
\begin{split}
& \dfrac{\Pi(\omega)}{\theta(\omega)} \overset{\text{exact}}{=} -\gamma (\mathbb{I}{-}i\omega \tau)^{-1}\kappa  =-\gamma \kappa-i\gamma \tau \kappa \, \omega +\mathcal{O}(\omega^2) \, , \\
& \dfrac{\Pi(\omega)}{\theta(\omega)} \, \, \, \overset{\text{IS}}{=} \, \, \, -\zeta(1{-}i\omega \tau_\Pi)^{-1}  = -\zeta -i\zeta \tau_\Pi \, \omega +\mathcal{O}(\omega^2) \, . \\
\end{split}
\end{equation}
This leads us to the following identifications, which are mandatory for interpretation (ii) to hold:
\begin{equation}\label{MagicIRED}
\begin{split}
\zeta ={}& \gamma \kappa = \sum_n \zeta_n \, , \\
\tau_\Pi ={}& \dfrac{\gamma \tau \kappa}{\gamma \kappa} = \dfrac{\sum_n \zeta_n \tau_n}{\sum_m \zeta_m} \, . \\
\end{split}
\end{equation}
As it turns out, these are the expressions of the transport coefficients within the IReD approach \cite{WagnerIReD2022}. This fact will be shown explicitly in Section \ref{IRUZZ}. However, it is evident a priori, since IReD fulfills interpretation (ii) by construction. 

Let us now focus on interpretation (iii). Here, the assumption is that, even if the Green function \eqref{Ghisa} cannot coincide with \eqref{Green!}, still one can look for the value of $\tau_\Pi$ that gives ``the best fit'' ($\zeta$ is fixed to $\gamma \kappa$ by the Navier-Stokes limit). In the original DNMR approach \cite{Denicol_Relaxation_2011,Denicol2012Boltzmann}, it is argued that, since the term in \eqref{Green!} with the largest relaxation time is the last to decay to zero, one should set $\tau_\Pi=\max\{\tau_n\}$. However, this approach has a serious limitation. In fact, suppose that the largest relaxation time, say $\tau_1$, has a very small associated susceptibility $\zeta_1$. Then, its contribution to the Green's function \eqref{Green!} may be negligible for all practical purposes. In this case, the DNMR prescription may dramatically overestimate the timescale over which $G$ relaxes to zero. The most natural way to avoid this problem is to define $\tau_\Pi$ as the average of the relaxation times $\tau_n$, weighted over their susceptibility $\zeta_n$. In this way, we give more importance to those relaxation times that contribute more to the Green function \eqref{Green!}. Quite surprisingly, this leads us back to the IReD prescription \eqref{MagicIRED}.

To appreciate these subtleties, let us consider a concrete example. Suppose that the ``exact model'' possesses three relaxation times, $\tau_n=\{5,2,1\}$, with associated susceptibilities $\zeta_n=\{1,1,8 \}$. Then, for $t>0$, the exact Green function \eqref{Green!}, the DNMR Green function \eqref{Ghisa}, and its IReD variation are respectively
\begin{equation}\label{greenish}
\begin{split}
 G_{\mathrm{ex}}(t) ={}& \dfrac{e^{-t/5}}{5} +\dfrac{e^{-t/2}}{2} +8 \, e^{-t} \, , \\
 G_{\mathrm{D}}(t) ={}& 2 \, e^{-t/5} \, , \\
 G_{\mathrm{IR}}(t) ={}& \dfrac{20}{3} e^{-2t/3} \, . \\
\end{split}
\end{equation}
\begin{figure}
\centering
\includegraphics{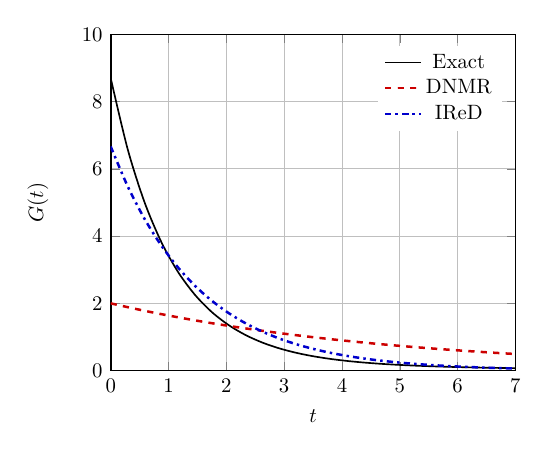}
\caption{\label{figuz1} Israel-Stewart approximations (dashed and dashdotted lines) for the linear-response Green function of a viscous flux (solid line) in a system with three relaxation times. The analytic formulas are provided in equation \eqref{greenish}.}
\end{figure}
While it is evident that DNMR fails to reproduce the transient dynamics of the system  (see Fig. \ref{figuz1}), IReD still provides a reasonable fit. Now, it should be emphasized that, since the Green function describes the response of $\Pi$ to an unrealistic instantaneous driving force $\theta(t)=\delta(t)$, in Fig. \ref{figuz1} the discrepancies between the models are exacerbated. In a situation where $\theta(t)$ varies over timescales that are comparable to the relaxation times, the agreement between the exact model, IReD, and DNMR can be surprisingly good, as we will show in our numerical study.

\section{Outline of the numerical study}

\begin{figure}
\centering
\includegraphics{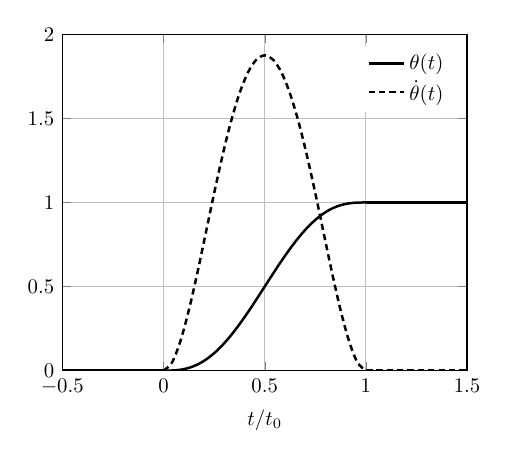}
\caption{The smoothstep function that is used as an external source to drive the internal degrees of freedom $\mathbb{A}$ (and thus the viscous flux $\Pi$) out of equilibrium. Its analytic profile can be found in equation \eqref{smoothy}.}
\label{fig:smooth}
\end{figure}

We consider a fluid element that is initially in thermodynamic equilibrium ($\mathbb{A}{=}0$), with no spatial gradients in the surroundings ($\theta{=}0$). Then, a perturbation comes along, and the driving force $\theta(t)$ is turned on, acting as an external source for $\Pi$ (in the linear response regime). To examine both the transient dynamics [interpretation (iii)] and the relaxation to the Navier-Stokes state [interpretations (i) and (ii)], we choose $\theta(t)$ to be a smoothstep that reaches 1 over a timescale comparable to the relaxation times. Its analytic profile is provided below\footnote{We would like to remark that our main results do not depend on the choice of $\theta(t)$. We tested many possible shapes, obtaining similar accuracy estimates. We chose the smoothstep because it illustrates all the essential physics quite well.}:
\begin{equation}\label{smoothy}
    \theta(t)=
\begin{cases}
      0 & \text{if } \,  t \leq 0 \, , \\
      6(t/t_0)^5{-}15(t/t_0)^4{+}10(t/t_0)^3        & \text{if } \, 0<t<t_0 \, , \\
      1        & \text{if } \, t \geq t_0 \, , \\
\end{cases}
\end{equation}
where $t_0$ is a parameter of the order of $\max\{ \tau_n \}$. The profile is visualized in figure \ref{fig:smooth}. We will vary the exact value of $t_0$ to explore more or less extreme transient regimes. Note that, although \eqref{smoothy} is not truly smooth, it is still of class $\mathcal{C}^2(\mathbb{R})$, which is good enough for our purposes.
Our goal is to ``rank'' different formulations of Israel-Stewart theory, based on their accuracy in reproducing the exact model prediction. Interpretation (i) will be considered outdone only if the Israel-Stewart prediction for $\Pi(t)$ is \textit{significantly} (i.e., visibly) more accurate than the Navier-Stokes prediction. 

In practice, we will compare six alternative dynamical equations for $\Pi$, which are supposed to approximate the exact model \eqref{SuperModel}. In all equations, the function $\theta(t)$ plays the role of an external source, and the transport coefficients are background constants.
\begin{flalign}
\text{DNMR:} && \label{two} \tau_{\text{D}} \dot{\Pi}+\Pi &= -\zeta \theta -\chi_{\text{D}} \dot{\theta} \;,&\\
\text{tDNMR:} && \label{three} \tau_{\text{D}} \dot{\Pi}+\Pi &= -\zeta \theta \;,&\\
\text{IReD:} && \label{four} \tau_{\text{IR}} \dot{\Pi}+\Pi &= -\zeta \theta \;,&\\
\text{1AA:} && \label{five} \tau_{(1)} \dot{\Pi}+\Pi &= -\zeta_{(1)} \theta \;,&\\
\text{2$^{\text{nd}}$OH:} && \label{six} \Pi &= -\zeta \theta- \chi_{2}\dot{\theta}  \;,&\\
\text{NS:} && \label{seven} \Pi &= -\zeta \theta \;.
\end{flalign}
 Equation \eqref{two} is the formulation of Israel-Stewart constructed by applying the original DNMR truncation procedure \cite{Denicol2012Boltzmann} to the ``exact'' model \eqref{SuperModel}. Note that there is an additional term $\propto \dot{\theta}$, which does not appear in the general Israel-Stewart theory \eqref{LowerModel}. This term is the explicit second-order gradient ($\mathcal{K}-$type \cite{Denicol2012Boltzmann}) contribution that makes DNMR acausal. Equation \eqref{three} is the truncated DNMR model that we obtain when we artificially remove this acausal term, as is always done in simulations. Equation \eqref{four} is constructed by applying the IReD approach \cite{WagnerIReD2022} to truncate the dynamics of \eqref{SuperModel}, reducing it again to a relaxation equation for $\Pi$ (with a different relaxation time with respect to DNMR). Equation \eqref{five} is the analogue of the 14-moment approximation (we call it ``1-affinity approximation''): it expresses the dynamics using the variables $\{\Pi,\mathbb{A}^2, \mathbb{A}^3,\ldots\}$, and then simply sets $\mathbb{A}^2{=}\mathbb{A}^3{=}\ldots{=}0$. Equation \eqref{six} is second-order hydrodynamics in its purest form: $\Pi$ is expanded up to second-order in gradients of the flow velocity $u^\mu$ (recall that $\theta$ is a first-order term). It is not properly an Israel-Stewart model, but it is instructive to use it as a reference, since within interpretation (ii) it should be as accurate as \eqref{two} and \eqref{four}, see Ref. \cite{Baier2008}. Finally, equation \eqref{seven} is the Navier-Stokes limit, which is used as our fundamental accuracy threshold. If an Israel-Stewart model fails to be appreciably more accurate than Navier-Stokes, then interpretation (i) applies to it (at best). 

In this section, we will derive equations \eqref{two}--\eqref{seven} one by one directly, as approximations of the ``exact'' model \eqref{SuperModel}, and we will express all transport coefficients in terms of the kinetic coefficients $\tau$, $\kappa$, $\gamma$.

\subsection{The Navier-Stokes limit}

Let us first show that the slow limit of model \eqref{SuperModel} is the Navier-Stokes theory. If we perform the change of integration variable $\Tilde{t}=t{-}t'$ in the second equation of \eqref{SuperPrediction}, we obtain the following (exact) formula for the viscous flux:
\begin{equation}
    \Pi(t) = -\int_0^{\infty} \gamma e^{-\tau^{-1} \Tilde{t}}\tau^{-1} \kappa \, \theta(t{-}\Tilde{t}) \d\Tilde{t} \, .
\end{equation}
Let us note that, if $\Tilde{t}\gtrsim \max\{\tau_n\}$, then the integrand is exponentially suppressed. Hence, the relevant contributions to the integral come from the region $0<\Tilde{t} \lesssim \max\{\tau_n\}$. This implies that, if the driving force $\theta$ varies slowly over the longest relaxation timescale, we can make the approximation $\theta(t{-}\Tilde{t})\approx \theta(t)$, and we can bring the force outside the integral. Then, using the well-known result of matrix calculus
\begin{equation}
    \int_0^{\infty} \! e^{-\tau^{-1} \Tilde{t}}\, \d\Tilde{t} = \tau \, ,
\end{equation}
we indeed recover the Navier-Stokes equation \eqref{seven}, with viscosity coefficient $\zeta=\gamma \kappa=\sum_n \zeta_n$.

\subsection{The DNMR truncation}

If we left-multiply the first equation of \eqref{SuperModel} by the row-vector $-\gamma$, we obtain
\begin{equation}\label{contruonzuz}
  -\gamma \tau \dot{\mathbb{A}}+\Pi =-\zeta \theta \, .  
\end{equation}
This tells us that the exact value of $\Pi$ can be expressed through the Navier-Stokes prescription, plus a relaxation-type correction that involves all the non-equilibrium degrees of freedom (in our case, all affinities $\mathbb{A}$). The main idea of DNMR is to keep the dynamics of the slowest eigenmode of the system, say, $\tau_1$, while the faster ones are approximated by their Navier-Stokes values \cite{Denicol2012Boltzmann}. To this end, we split the first term of \eqref{contruonzuz} into two pieces as follows:
\begin{equation}
   -\gamma \tau \dot{\mathbb{A}} = \tau_1 \dot{\Pi} + \gamma(\tau_1\mathbb{I}{-}\tau)\dot{\mathbb{A}} \, .   
\end{equation}
Then, we define the eigenprojected vectors $\mathbb{X}_n=\mathcal{P}_n \mathbb{A}$, so that we have $\mathbb{A}=\sum_n \mathbb{X}_n$ (because $\sum_n \mathcal{P}_n = \mathbb{I}$). This allows us to rewrite \eqref{contruonzuz} as follows:
\begin{equation}\label{bresco}
    \tau_1 \dot{\Pi}+\Pi = -\zeta \theta -\gamma \sum_{n \neq 1} (\tau_1{-}\tau_n)\dot{\mathbb{X}}_n \, .
\end{equation}
Up to this point, no approximation has been made (we removed the $n{=}1$ contribution to the sum because it vanishes). Now, we notice that, if we left-multiply the first equation of \eqref{SuperModel} by $\mathcal{P}_n$, we obtain $\tau_n\dot{\mathbb{X}}_n+\mathbb{X}_n=\mathcal{P}_n \kappa \theta$, which, when differentiated in time, becomes
\begin{equation}\label{stuty}
    \dot{\mathbb{X}}_n = \mathcal{P}_n \kappa \dot{\theta} + \mathcal{O}(\tau_n \partial^3) \, .
\end{equation}
The DNMR truncation consists of using equation \eqref{stuty} to replace $\dot{\mathbb{X}}_n$ in equation \eqref{bresco}. In this way, consistency with interpretation (ii) is preserved, because the error is of order 3 in gradients (recall that $\mathbb{X}_n$ is already of order 1). Furthermore, also interpretation (iii) is somehow respected, because the sum in equation \eqref{bresco} does not involve the $n=1$ term, meaning that the transient dynamics of $\mathbb{X}_1$ is captured exactly. The result is the DNMR equation of motion \eqref{two}, with transport coefficients
\begin{equation}\label{eqs:coeff_DNMR}
\begin{split}
\tau_{\text{D}} &=  \tau_1 \;, \\
\chi_{\text{D}} &=  \gamma(\tau_1{-}\tau)\kappa= \zeta \left(\tau_1-\frac{\gamma\tau \kappa}{\gamma\kappa}\right) \;.
\end{split}
\end{equation}

\subsection{The IReD truncation and second-order hydrodynamics}\label{IRUZZ}

Both the IReD approach \cite{WagnerIReD2022}, and second-order hydrodynamics \cite{Baier2008} adopt interpretation (ii), namely that Israel-Stewart theories are refinements of Navier-Stokes, and should be accurate up to second order in the derivatives. In fact, both of them start from the assumption that the process is very slow, so that equations \eqref{SuperModel} and \eqref{contruonzuz} imply
\begin{equation}
    \begin{split}
\mathbb{A} ={}& \kappa \theta +\mathcal{O}(\partial^2) \, , \\
\Pi ={}& -\zeta \theta +\mathcal{O}(\partial^2) \, . \\
    \end{split}
\end{equation}
These equations, by themselves, are not accurate enough for our purposes, because their error is of order 2 in derivatives. However, if we differentiate them with respect to time, we get two possible approximations for $\dot{\mathbb{A}}$, namely
\begin{equation}
    \begin{split}
\dot{\mathbb{A}}={}& -\kappa \dot{\Pi}/\zeta +\mathcal{O}(\partial^3) \, , \\
\dot{\mathbb{A}}={}& \kappa \dot{\theta} +\mathcal{O}(\partial^3) \, . \\
    \end{split}
\end{equation}
If we use the first approximation to replace $\dot{\mathbb{A}}$ in equation \eqref{contruonzuz}, we recover the IReD prescription, see equation \eqref{four}, whereas, if we use the second approximation for $\dot{\mathbb{A}}$, we recover second-order hydrodynamics, as given in equation \eqref{six}. The associated transport coefficients are (recall that $\zeta=\gamma \kappa$) 
\begin{equation}
    \tau_{\mathrm{IR}} = -\dfrac{\chi_2}{\zeta} = \dfrac{\gamma \tau \kappa}{\gamma \kappa} = \dfrac{\sum_n \zeta_n \tau_n}{\sum_m \zeta_m} \, .
\end{equation}
Let us make an important remark. Both IReD and second-order hydrodynamics arise from the same underlying assumption, and adopt the same type of approximation. Hence, their accuracy is necessarily similar for very slow processes. However, when $\theta(t)$ varies over timescales that are comparable to the relaxation times, the two equations are \textit{very different}. In fact, the linear-response Green function of second-order hydrodynamics is $G_2(t){=}\zeta \delta(t){+}\chi_2 \delta'(t)$, which has nothing to do with the exact Green function \eqref{Green!}. Thus, the transient evolution of second-order hydrodynamics is expected to be the least accurate, among equations \eqref{two}-\eqref{six}. On the other hand, even if the IReD approach assumes interpretation (ii), the value of the IReD relaxation time happens to be the weighted average \eqref{MagicIRED}. As a result, the IReD Green function turns out to be a very accurate exponential approximation of the exact Green function.

\subsection{The 1-affinity approximation}
\label{sec:1AA}
The 1-affinity approximation is the abstract generalization of Grad's 14-moment approximation \cite{Israel_Stewart_1979}. It differs from all others in the fact that it never assumes that the dynamics is ``slow''. Instead, it uses the second equation of \eqref{SuperModel} to rewrite the equation of motion for $\A^1$ as an equation of motion for $\Pi$:
\begin{equation}
    \tau\indices{^1 _1} \dot{\Pi}+\sum_{n \neq 1} (\tau\indices{^1 _1} \gamma_n -\gamma_1 \tau\indices{^1 _n})\dot{\A}^n+\Pi+\sum_{n\neq 1} \gamma_n \A^n=-\gamma_1 \kappa_1 \theta \;.
\end{equation}
Neglecting the impact of all other degrees of freedom completely, by just setting $\A_n=0$ for $n\neq 1$, we obtain a closed equation for the viscous stress. This has the form \eqref{five}, with
\begin{equation}
    \tau_{(1)}=\tau^1{}_{1}\;, \quad \quad \quad  
    \zeta_{(1)}=\gamma_1\kappa_1\;.
\end{equation}
We remark that this approximation has to be slightly modified in the case of kinetic theory discussed in section \ref{sec:kin}, where the fundamental matrix is given by $\tau^{-1}$ (and not $\tau$ itself). 

\section{Numerical results}
In this section, we present our numerical results on the performance of the different Israel-Stewart-type prescriptions introduced in the previous sections.

\subsection{Two degrees of freedom}

To assess the performance of the different theories introduced in the preceding section, we first consider the simple case of two affinities, i.e., $\mathbb{A}=\{\mathbb{A}^1,\mathbb{A}^2\}^{\mathrm{T}}$. This case is admittedly quite far from the kinetic theory problem, where the non-equilibrium degrees of freedom are infinite. However, it may be important for applications to bulk-viscous neutron stars \cite{Camelio2022,GavassinoBurgers2023}. In practice, to completely specify the ``exact'' model, we only need to prescribe the eigenvalues $\tau_n$ and susceptibilities $\zeta_n$, since this fully determines the exact Green function. Indeed, most of the Israel-Stewart prescriptions for the transport coefficients discussed above can be expressed only in terms of $\tau_n$ and $\zeta_n$. The only exception is the one-affinity approximation, which requires the  knowledge of $\tau$, $\kappa$, and $\gamma$. For simplicity, we will assume that $\tau=\text{diag}(\tau_1,\tau_2)$, so that the one-affinity transport coefficients simply become $\tau_{(1)}=\tau_1$ and $\zeta_{(1)}=\zeta_1$. 

\subsubsection{Varying the susceptibilities}
In order to explore the regimes where we expect different hydrodynamic theories to be valid, we start by varying the susceptibilities $\zeta_n=\{\zeta_1,\zeta_2\}$ that determine how important each eigenvalue is for the evolution of $\Pi$.
Specifically, let us consider the following parameter sets:
\begin{equation*}
\begin{split}
    \mathrm{(a)}\;\; \tau_n=\{5,1\}\;, & \;\zeta_n=\{5,1\}\;,\; t_0=5\;, \\
    \mathrm{(b)}\;\; \tau_n=\{5,1\}\;, & \;\zeta_n=\{1,1\}\;,\;t_0=5\;, \\
    \mathrm{(c)}\;\; \tau_n=\{5,1\}\;, & \;\zeta_n=\{1,5\}\;,\;t_0=5\;, \\
    \end{split}
\end{equation*}
where $t_0$ is the parameter appearing in equation \eqref{smoothy}. Let us discuss these scenarios one by one.

\begin{figure}[ht]
    \centering
    \includegraphics[width=.49\textwidth]{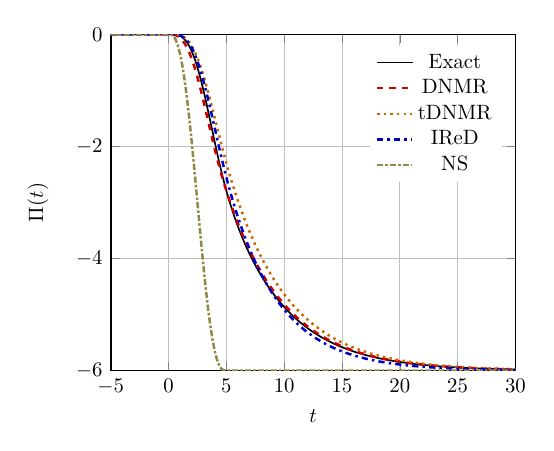}
    \includegraphics[width=.49\textwidth]{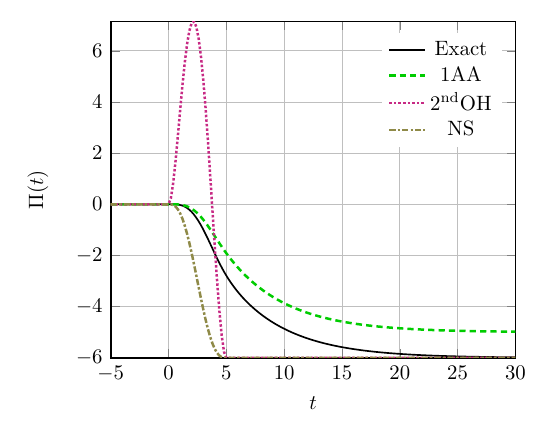}
    \caption{The time evolution of $\Pi(t)$ for parameter set (a) in different hydrodynamic approaches compared to the exact results. The black line, representing the exact result, is the same in both pictures.}
    \label{fig:plots_a}
\end{figure}

In case (a), the slowest scale is also the most important one to $\Pi$ since $\zeta_1$ is considerably larger than $\zeta_2$. As is evident from figure \ref{fig:plots_a}, both DNMR and IReD capture the dynamics very well. Indeed, the error of DNMR drops off faster than the one of IReD, especially at late times. This is to be expected, since the parameter set (a) lies precisely in the regime where the assumptions of DNMR hold. However, when omitting the term $\sim \dot{\theta}$, i.e., considering the tDNMR method, the agreement with the exact solution, while still reasonably good, becomes worse than the one of IReD. In comparison to the previous three theories, the performances of 1AA, 2$^{\mathrm{nd}}$OH, and NS, are considerably worse, cf. figure \ref{fig:plots_a}. While 1AA does not capture the correct late-time limit, NS relaxes to it on a timescale that is much too short. Finally, 2$^{\mathrm{nd}}$OH, in addition to relaxing too fast to the late-time limit, features an unphysical upward spike that is induced by the nonvanishing derivative $\dot{\theta}$. Only at very late times $t_{\mathrm{NS}}\sim 5 \max\{\tau_n\}$, the different methods (excluding 1AA) agree.
    
\begin{figure}[ht]
    \centering
    \includegraphics[width=.49\textwidth]{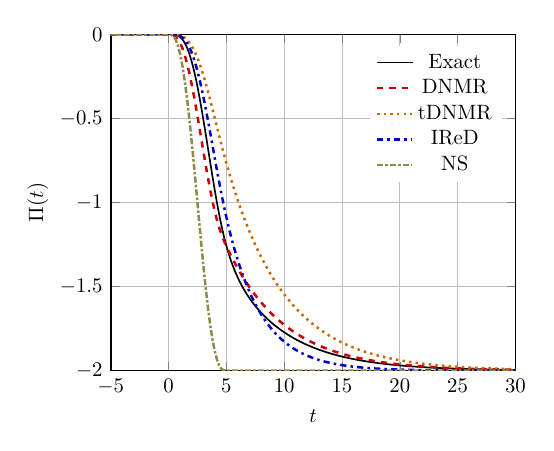}
    \includegraphics[width=.49\textwidth]{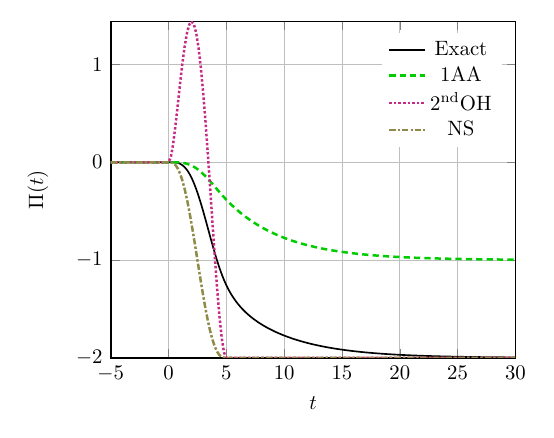}
    \caption{Same as figure \ref{fig:plots_a}, but for parameter set (b).}
    \label{fig:plots_b}
\end{figure}

    In parameter set (b), both the slow and the fast mode are equally important. Considering the results displayed in figure \ref{fig:plots_b}, both DNMR and IReD perform rather well, with DNMR converging to the exact solution faster at late times. However, in this case the omission of the term $\sim\dot{\theta}$ has a larger effect on the performance of tDNMR, making it visibly worse than IReD. The qualitative behavior of the 1AA, 2$^{\mathrm{nd}}$OH and NS is the same as in case (a), with 1AA incurring a greater error in the Navier-Stokes value. 
    
    Parameter set (c) inverts the weight that each mode contributes to the quantity $\Pi$, with the faster mode being more important than the slower one. In accordance with the previous discussions, DNMR, while still catching some features of the exact curve, has trouble dealing with the importance of the slower mode. This problem is worsened in tDNMR, which now is significantly off. IReD, on the other hand, does not suffer considerably, and captures the transient dynamics relatively well. Furthermore, because the faster mode now has a greater influence, $\Pi$ reaches its asymptotic limit faster, making both NS and 2$^{\mathrm{nd}}$OH perform better in comparison to cases (a) and (b).

\begin{figure}[ht]
    \centering
    \includegraphics[width=.49\textwidth]{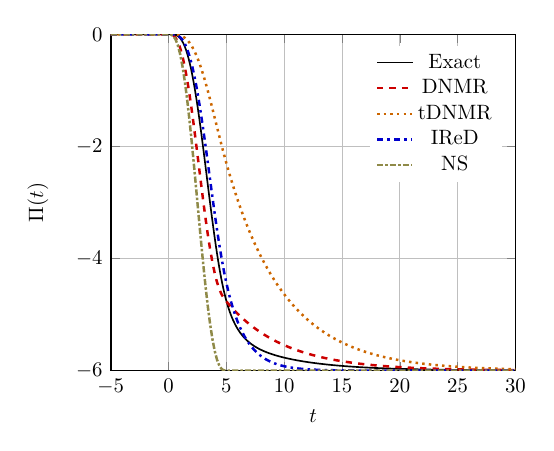}
    \includegraphics[width=.49\textwidth]{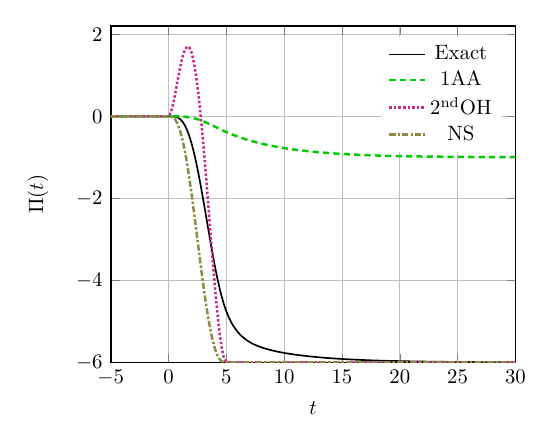}
    \caption{Same as figure \ref{fig:plots_a}, but for parameter set (c).}
    \label{fig:plots_c}
\end{figure}

\subsubsection{Varying the transient regime}
\begin{figure}[ht]
    \centering
    \includegraphics[width=.49\textwidth]{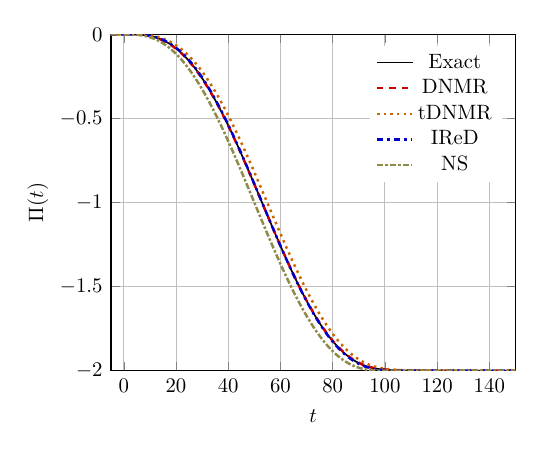}
    \includegraphics[width=.49\textwidth]{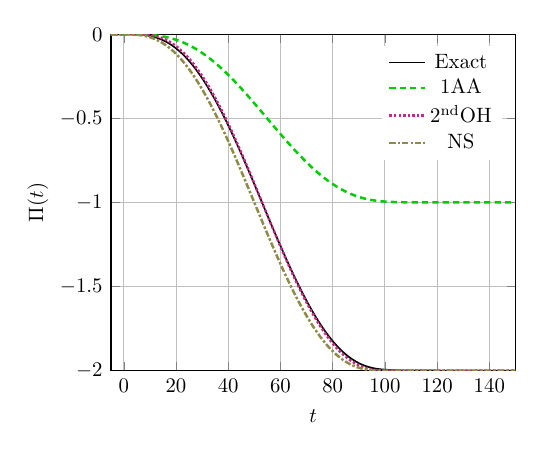}
    \caption{Same as figure \ref{fig:plots_a}, but for parameter set (d).}
    \label{fig:plots_d}
\end{figure}

In the previous subsubsection we considered a smoothstep that varied on a timescale equal to that of the longest microscopic relaxation time. To test more or less violent transient regimes, we consider the following additional parameter sets,
\begin{equation*}
\begin{split}
    \mathrm{(d)}\;\; & \tau_n=\{5,1\}\;, \;\zeta_n=\{1,1\}\;,\; t_0=100\;,  \\\
    \mathrm{(e)}\;\; & \tau_n=\{5,1\}\;, \;\zeta_n=\{1,1\}\;,\;t_0=0.01\;. \\
    \end{split}
\end{equation*}
Intuitively, one would expect the performances of NS and 2$^{\mathrm{nd}}$OH to become worse as the timescale $t_0$ is decreased. Indeed, at the very slow dynamics of case (d) depicted in figure \ref{fig:plots_d}, essentially all theories (with the exception of 1AA) capture the evolution of the system correctly, although DNMR and IReD are still the most accurate.
Case (e) is opposite to the previous one, since $t_0$ is now negligible and the source term $\theta$ is basically a step function. As expected, the NS and 2$^{\mathrm{nd}}$OH prescriptions fail to describe the evolution of $\Pi$ in the transient regime, while the error of DNMR, tDNMR, and IReD is increased. Interestingly, in the case of DNMR, one can see the effect of the $\dot{\theta}$-term in a jump at $t{=}0$ that does not coincide with the exact evolution.

\begin{figure}[ht]
    \centering
    \includegraphics[width=.49\textwidth]{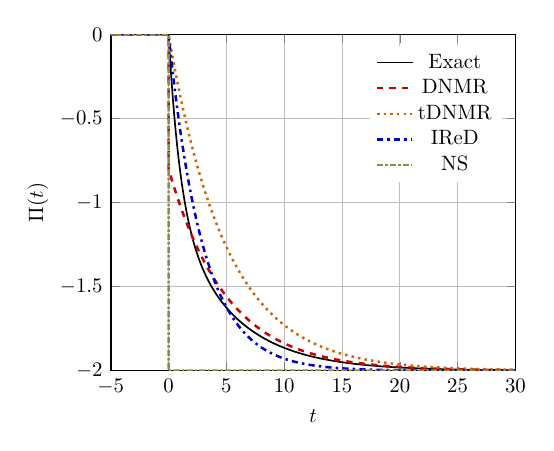}
    \includegraphics[width=.49\textwidth]{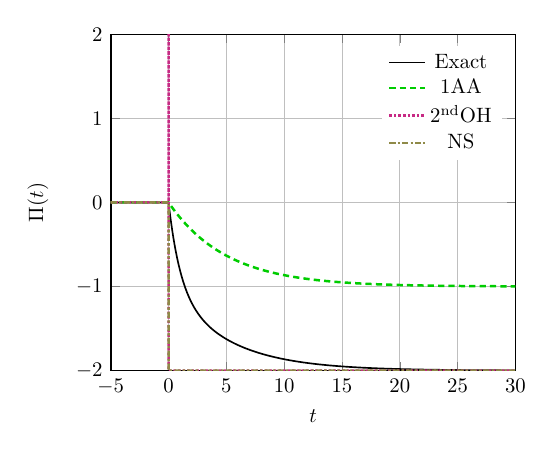}
    \caption{Same as figure \ref{fig:plots_a}, but for parameter set (e).}
    \label{fig:plots_e}
\end{figure}

\subsubsection{Ranking the theories}
One can rank the performance of the different theories by comparing how close they are to the exact solution, which we do by considering the square root of the integrated square deviations weighed by the asymptotic magnitude of $\Pi$,
\begin{equation}
  D= \frac{1}{\zeta}\sqrt{\int_{\mathbb{R}} \left[\Pi_{\mathrm{exact}}(t')-\Pi(t')\right]^2  \d t'}   \;.\label{eq:def_D}
\end{equation}
When excluding 1AA, which does not capture the Navier-Stokes value correctly, such a procedure can assess how well the different theories describe the transient dynamics. The results are listed in table \ref{tbl:scores}. Schematically, we find the following orderings,
\begin{align*}
    \text{(a),(b): DNMR $\succ$ IReD $\succ$ tDNMR $\succ$ NS $\succ$ 2$^{\mathrm{nd}}$OH}\;,\qquad \quad
    \text{(c): IReD $\succ$ DNMR $\succ$ 2$^{\mathrm{nd}}$OH $\succ$ NS $\succ$ tDNMR}\;,\\
    \text{(d): DNMR $\succ$ IReD $\succ$ 2$^{\mathrm{nd}}$OH $\succ$ tDNMR $\succ$ NS }\;,\qquad \quad
    \text{(e): IReD $\succ$ DNMR $\succ$ tDNMR $\succ$ NS $\succ$  2$^{\mathrm{nd}}$OH}\;.
\end{align*}
where $x\succ y$ means that method $x$ performs better than method $y$.
It should be noted that while DNMR performs best in its regime of validity, IReD should be preferred in practice, since it gives better results than the (actually used) tDNMR method. In the regime where the fast modes contribute greatly to the dynamics of $\Pi$, the tDNMR method that is implemented in practice actually performs worst, while IReD still gives good results. We remark that the worsening of the performance of tDNMR are in agreement with Ref. \cite{Denicol:2012vq}, where the heat-flow problem was considered for a system described by kinetic theory. In fact, the resummed ``21/37''-theory of fluid dynamics proposed there to restore agreement with the exact results, although introducing more dynamical degrees of freedom, shares some important features with IReD, cf. Sec. V of Ref. \cite{WagnerIReD2022}.

\begin{table}[ht]
    \centering
    \begin{tabular}{|c||c|c|c|c|c|}
         \hline& DNMR & tDNMR & IReD  & 2$^{\mathrm{nd}}$OH & NS  \\ \hline\hline
         (a)&0.050   &0.18   & 0.082  & 1.92 & 1.25 \\\hline
         (b)&0.15   &0.55   & 0.18   & 1.23  & 0.91\\\hline
         (c)&0.25   &0.92  & 0.13   & 0.56 & 0.59 \\\hline
         (d)&0.0080   &0.24  & 0.016  & 0.052  & 0.35 \\\hline
         (e)&0.27   &0.58  & 0.20   & 35.83  & 1.09 \\\hline
    \end{tabular}
    \caption{The error $D$, defined in equation \eqref{eq:def_D}, for different methods and parameter sets (a)--(e). In cases (a), (b), (c), and (e), we cut off the range of numerical integration at $t_{\mathrm{max}}=50$, while we use $t_{\mathrm{max}}=300$ in case (d).}
    \label{tbl:scores}
\end{table}

\subsection{300 random degrees of freedom}
\begin{figure}[ht]
    \centering
    \includegraphics[width=.49\textwidth]{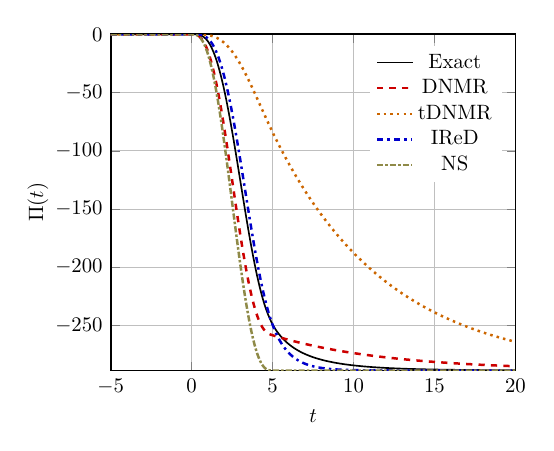}
    \includegraphics[width=.49\textwidth]{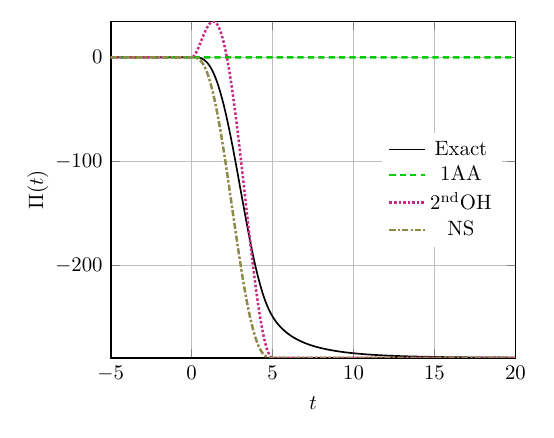}
    \caption{The time evolution of $\Pi(t)$ in different hydrodynamic approaches for a system of 300 independent degrees of freedom with exponentially distributed relaxation times and susceptibilities.}
    \label{fig:plots_300}
\end{figure}

In this subsection, we consider a system with many (300) degrees of freedom which relax on different timescales. Both the eigenvalues $\tau_n =\{\tau_1,\ldots,\tau_{300}\}$, and the associated susceptibilities, $\zeta_n=\{\zeta_1,\ldots,\zeta_{300}\}$, are independently drawn from an exponential distribution with unit mean and variance, i.e.,
\begin{equation}
    P(\tau_i\leq x)=P(\zeta_i\leq x)=\begin{cases}
        1-e^{-x}\;,\;\; &x\geq 0 \\
        0\;,\;\;&x<0
    \end{cases}\qquad \forall\, i\,\in\, \{1,\ldots,300\}\;.
\end{equation}
We again use the smoothstep \eqref{smoothy} as a source, with growth-rate parameter $t_0=5$, which is five times larger than the mean relaxation time $\text{E}(\tau_n)=1$, but of the order of the expected maximal relaxation time $E(\max\{\tau_n\})=H_{300}\approx 6.3$, where $H_n$ denotes the $n$-th harmonic number. From figure \ref{fig:plots_300}, it is immediately clear that the timescale of the dynamics is too short for NS or 2$^{\mathrm{nd}}$OH to describe them. Similarly, 1AA does not catch the correct Navier-Stokes value or the dynamics at all, since it completely neglects 299 contributions. DNMR on the other hand qualitatively captures the dynamics, but does so in a worse way than IReD, which agrees rather well with the exact solution. Finally, tDNMR is also off by a large amount, performing even worse than NS and 2$^{\mathrm{nd}}$OH. The ranking of the different theories is very clear:
\begin{equation*}
    \text{IReD $\succ$ DNMR $\succ$ 2$^{\mathrm{nd}}$OH $\succ$ NS $\succ$ tDNMR}\;.
\end{equation*}
The case of randomly assigned susceptibilities and associated relaxation times is of course a rather extreme one; however, it clearly shows the difficulties that hydrodynamic theories like NS face when they are applied to a system with many degrees of freedom which are important. The refined second-order approaches DNMR and IReD on the other hand are able to surmount these challenges and still provide a rather good description of the system at all times. As in the previous section, however, it has to be stressed that the truncated tDNMR used in practice does not work well in this regard.

\subsection{A real example from kinetic theory}
\label{sec:kin}
\begin{figure}[ht]
    \centering
    \includegraphics[width=.49\textwidth]{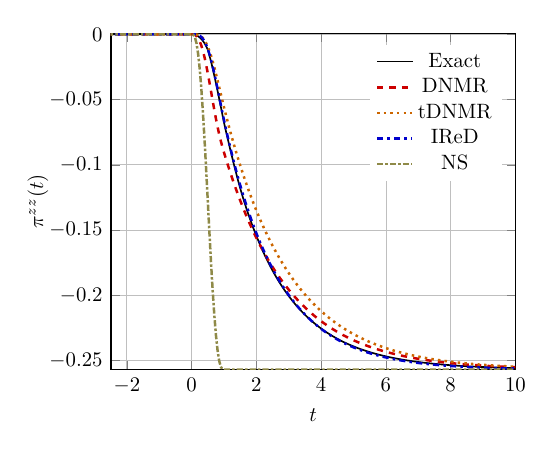}
    \includegraphics[width=.49\textwidth]{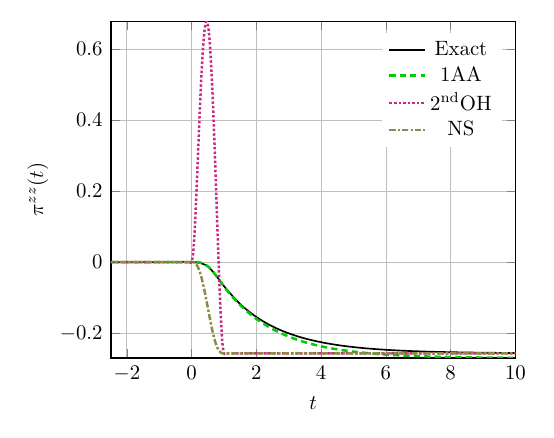}
    \caption{The time evolution of $\pi^{zz}(t)$ in different hydrodynamic approaches for a coupled system of 50 moments, describing a system of particles subject to hard-sphere collisions.}
    \label{fig:plots_kin}
\end{figure}

As a final test case, we consider kinetic theory for a gas of ultrarelativistic particles interacting via a constant cross section. For the basic concepts of relativistic kinetic theory, we refer the reader to {Appendix \ref{app:kin_theory} and} Refs. \cite{DeGroot:1980dk,Cercignani}. In the context of this {section}, it is sufficient to state that the equations of motion for the so-called irreducible moments of tensor-rank two\footnote{We choose to go with the moments of tensor-rank two (instead of zero) since the bulk viscous pressure vanishes in the ultrarelativistic limit $m/T\to 0$, where $m$ is the particle mass.} of the distribution function take on the following form \cite{Denicol2012Boltzmann},
\begin{equation}
    \dot{\rho}_r^{\mu\nu}+\sum_{n=0}^{N_2}\mathcal{A}_{rn}^{(2)}\rho_n^{\mu\nu}=\frac{(4+r)!}{15\pi^2} T^{4+r} \sigma^{\mu\nu}+\cdots\;,\label{eq:momeq}
\end{equation}
where the ellipses denote quadratic terms that we will not consider in our simplified setup, and $T$ is the temperature, which we take to be constant. Additionally, $N_2$ denotes the truncation order, i.e., the number of coupled equations we consider. Furthermore, $\sigma^{\mu\nu}=\partial^{\langle\mu}u^{\nu\rangle}$ is the shear tensor, $u^\mu$ is the four-velocity of the fluid, and the angular brackets denote the projection onto the subspace of tensors that are traceless, symmetric and orthogonal to $u^\mu$\footnote{Specifically, we have for a second-rank tensor $A^{\langle\mu\nu\rangle}=\Delta^{\mu\nu}_{\alpha\beta} A^{\alpha\beta}$, and $\Delta^{\mu\nu}_{\alpha\beta}=\frac12(\Delta^\mu_\alpha\Delta^\nu_\beta+\Delta^\mu_\beta\Delta^\nu_\alpha)-\frac13 \Delta^{\mu\nu}\Delta_{\alpha\beta}$, with the basic projector $\Delta^{\mu\nu}=g^{\mu\nu}-u^\mu u^\nu$.}.
The linearized collision matrix $\mathcal{A}^{(2)}$ has been recently computed exactly in Ref. \cite{Wagner:2023joq}, and is given by
\begin{equation}
    \mathcal{A}^{(2)}_{0n}= \frac{432 (-T)^{-n}}{\lambda_{\text{mfp}}(n+5)!}S_n^{(2)}(N_2)\;,\quad \mathcal{A}^{(2)}_{r>0,n\leq r} =\frac{T^{r-n}(r+4)!(n+1)}{\lambda_{\text{mfp}}(n+5)!r(r+1)}(9n+nr-4r)\left(\delta_{nr}-\frac{2}{r+2}\right)\;,\quad \mathcal{A}^{(2)}_{r>0,n> r}=0\;,
\end{equation}
where $\lambda_{\text{mfp}}$ is the mean free path, and we defined
\begin{equation}
    S_n^{(2)}(N_2)=\sum_{m=n}^{N_2} \binom{m}{n} \frac{1}{(m+2)(m+3)}\;.
\end{equation}
Since the entries of the matrix $\mathcal{A}^{(2)}$ are proportional to different powers of the temperature, $\mathcal{A}^{(2)}_{rn}\sim T^{r-n}$, it is sensible to define the de-dimensionalized moments $\widetilde{\rho}_r^{\,\mu\nu}=\rho_r^{\mu\nu}/T^r$, and the dimensionless matrix $\widetilde{\mathcal{A}}^{(2)}_{rn}=\mathcal{A}^{(2)}_{rn} T^{n-r}$. Then, we can rewrite equation \eqref{eq:momeq} as (choosing $\mu=\nu=z$)
\begin{equation}
    \sum_{n=0}^{N_2}\tau^{(2)}_{rn} \dot{\widetilde{\rho}}^{\,zz}_n +\widetilde{\rho}_r^{\,zz}= \eta_r T^4 \sigma^{zz}\;,\label{eq:eom_rhoz}
\end{equation}
where we defined 
\begin{equation}
    \tau^{(2)}=\left(\widetilde{\mathcal{A}}^{(2)}\right)^{-1} \qquad \mathrm{and}\qquad \eta_r= \sum_{n} \tau_{rn}^{(2)} \frac{(4+n)!}{15\pi^2}\;.
\end{equation}
Comparing equation \eqref{eq:eom_rhoz} and the first line of \eqref{SuperModel}, we can identify $\tau \equiv \tau^{(2)}$, $\kappa\equiv \eta$, and $\theta=T^4 \sigma^{zz}$. The quantity of interest from a hydrodynamic perspective is the $zz$-component of the shear-stress tensor $\pi^{zz}=\widetilde{\rho}_0^{\,zz}$, such that $\gamma=\{-1,0,0,\ldots\}$. The vector $\zeta_n$ can then be computed according to equation \eqref{Green!}.
Choosing $t_0=\lambda_{\text{mfp}}=1$, we simulate the performance of the hydrodynamic approximations to the exact dynamics of the system \eqref{eq:eom_rhoz} (considering the coupled dynamics of $N_2=50$ moments\footnote{We remark that the 1AA approximation (corresponding to the 14-moment approximation in kinetic theory) requires us to restrict the matrix $\tau$ to one entry, i.e., $\tau^{(2)}\equiv\tau^{(2)}_{00}=(\widetilde{\mathcal{A}}^{(2)}_{00})^{-1}$. This is because the procedure described in section \ref{sec:1AA} has to be carried out for the system \eqref{eq:momeq}, which features $\mathcal{A}^{(2)}$ and not its inverse.}) reacting to the smoothstep \eqref{smoothy} (that we multiply by $-1$ to facilitate a better visual comparison with the previous cases) which now can be interpreted as a shearing of the fluid. The results are displayed in figure \ref{fig:plots_kin}. 

The conclusions drawn from this realistic case do not change much from the simpler models discussed before:
Neither NS nor 2$^{\mathrm{nd}}$OH describe the transient dynamics correctly, and 1AA does not capture the late-time limit, although the error is significantly less than in the previous examples. On the other hand, the second-order theories are doing rather well, with IReD performing best, its predictions lying on top of the exact solution, which is in agreement with the results obtained for the relaxation-time approximation in Ref. \cite{Ambrus:2022vif}. The different formulations, whose errors are listed in table \ref{tbl:scores_kin}, can be ranked as follows,
\begin{equation*}
    \text{IReD $\succ$ DNMR $\succ$ tDNMR $\succ$ NS $\succ$ 2$^{\mathrm{nd}}$OH}\;.
\end{equation*}
Finally, it should be noted that the error of IReD is an order of magnitude lower than the ones of DNMR and tDNMR. This suggests that the complicated coupling between the moments may result in the slowest degree of freedom having a smaller susceptibility than the combination of the faster ones, thereby breaking the fundamental assumption of DNMR, in favor of the IReD approach, which is more agnostic. 

\begin{table}[ht]
    \centering
    \begin{tabular}{|c|c|c|c|c|c|}
         \hline DNMR & tDNMR & IReD  & 2$^{\mathrm{nd}}$OH & NS  \\ \hline\hline
         0.095   &0.13   & 0.013   & 1.79 & 0.84 \\\hline
    \end{tabular}
    \caption{The error $D$, defined in equation \eqref{eq:def_D}, for different methods in the case of ultrarelativistic kinetic theory.}
    \label{tbl:scores_kin}
\end{table}

\section{Some additional remarks}

We conclude our analysis with some useful observations, which should help the reader interpret our results.

\subsection{Breakdown of the gradient expansion}

Our numerical experiment clearly shows that second-order hydrodynamics (i.e. \textit{non-resummed} BRSSS \cite{Baier2008}) may be even less accurate than Navier-Stokes hydrodynamics, in the transient regime. This feels rather counterintuitive, since one would think that, in the grand scheme of the gradient expansion, adding more and more terms would result in better and better accuracy, until the exact solution is recovered. Unfortunately, this is not always true.

In order to see this, let us consider again the exact formula for the viscous flux computed from linear response theory: $\Pi(t)=-\int_{\mathbb{R}} G(t{-}t')\theta(t')\d t'$. We can express $\theta(t')$ as a Taylor series centered at $t$, so that we get
\begin{equation}
    \Pi(t) = -\sum_{k=0}^{\infty} \dfrac{(-1)^k}{k!}\dfrac{\d^k \theta(t)}{\d t^k} \int_{\mathbb{R}} G(s)s^k \d s \, .
\end{equation}
Recalling that the Green function is given by \eqref{Green!}, we can compute the integrals analytically, and we obtain
\begin{equation}\label{bruco}
    \Pi(t) = -\sum_n \zeta_n \sum_{k=0}^{\infty}(-\tau_n)^k \dfrac{\d^k \theta(t)}{\d t^k}  \, .
\end{equation}
This is the gradient expansion of the flux $\Pi(t)$, expressed in terms of the derivatives of the hydrodynamic variable $\theta(t)$. If we truncate at $k{=}0$, we recover Navier-Stokes, if we include the term $k{=}1$, we obtain second-order hydrodynamics, and so on. To understand the properties of the series \eqref{bruco}, let us assume that, in a neighborhood of $t$, the flow $\theta$ is exponentially relaxing, i.e. $\theta(t')=\theta_0 \, e^{-t'/t_0}$ for $t' \in (t{-}\varepsilon,t{+}\varepsilon)$, with some $\varepsilon>0$. Then, equation \eqref{bruco} reduces to
\begin{equation}
    \Pi(t)=-\theta(t)\sum_n \zeta_n \sum_{k=0}^{\infty} \bigg( \dfrac{\tau_n}{t_0}\bigg)^k \, .
\end{equation}
As can be seen, the series is geometric, and thus converges only if $t_0 > \tau_n$. This implies that, if $\theta(t)$ varies over timescales that are shorter than a relaxation time, the gradient expansion is divergent, and there is no guarantee that adding more and more terms will improve the accuracy. Instead, the more terms we add to the gradient expansion, the more we should expect to see large fluctuations in the transient regime, as we indeed see, e.g., in figure \ref{fig:plots_kin}.

There is also an additional problem. When we replaced $\theta(t)$ with its Taylor series, we made the implicit assumption that $\theta(t)$ was analytic. Indeed, when the perturbation is non-analytic (as is the case with our smooth-step), no gradient expansion is able to reproduce the correct value of $\Pi$, because local gradients cannot ``remember'' what the state was before a non-analyticity point. This introduces a large error in the predictions of second-order (and higher-order) hydrodynamics in proximity to all non-analyticity points. Such error decays over a timescale $\text{max}\{\tau_n\}$, meaning that, again, the whole transient evolution is modeled incorrectly. 

We can illustrate this issue with a simple example. Suppose that, for $t<0$, the fluid element is in thermodynamic equilibrium, so that $\Pi(0^-)=\theta(0^-)=0$. At $t=0$, a non-analytic perturbation arrives, with profile $\theta(t)=t$ for $t>0$. Then, from equation \eqref{SuperPrediction}, we find that the ``exact'' viscous stress is continuous at $t=0$, since
\begin{equation}
\Pi(0^+)={-} \lim_{t \rightarrow 0^+}\int_{0}^t \gamma e^{-\tau^{-1}(t-t')} \tau^{-1}\kappa t'\d t' =0 \, .
\end{equation}
Indeed, also the Navier-Stokes prediction, $\Pi(t)=- \Theta(t)\zeta t$, is continuous. However, if we go up to second order, there is a discontinuity: $\Pi(t)=-\Theta(t)(\chi_2+\zeta t)$. This is no surprise: The gradient expansion assumes that $\theta(t)$ is analytic, and therefore it ``behaves'' as if $\theta(t)$ were equal to $t$ also at negative times.

\subsection{Continuous spectrum: an analytic example}

For computational reasons, in our numerical study, we restricted our attention to systems with a finite number of degrees of freedom. However, since most of our results do not seem to depend on the number of affinities, we expect them to remain valid also in the limit of infinite affinities. Let us verify this explicitly with an exactly solvable model. 

In the infinite-dimensional case, the operator $\tau$ may possess a continuous spectrum, in which case \eqref{gigia} no longer reduces to \eqref{Green!}, but we need to consider its continuous analogue:
\begin{equation}\label{chervone}
    G(t)=\Theta(t) \int_{0}^{\infty} \zeta(\nu) \nu e^{-\nu t} \d\nu = -\Theta(t) \, \dfrac{\d}{\d t}\mathcal{L}\big\{\zeta(\nu) \big\}(t) \, ,
\end{equation}
where $\zeta(\nu)$ may be interpreted as the spectral density of susceptibility, $\nu$ is a continuous parameter spanning all possible values of [relaxation time]$^{-1}$, and $\mathcal{L}$ is the Laplace transform. The presence of the Laplace transform tells us that $G(t)$ no longer behaves as a sum of exponentials. For example, let us take
\begin{figure}
\centering
\includegraphics[width=.49\textwidth]{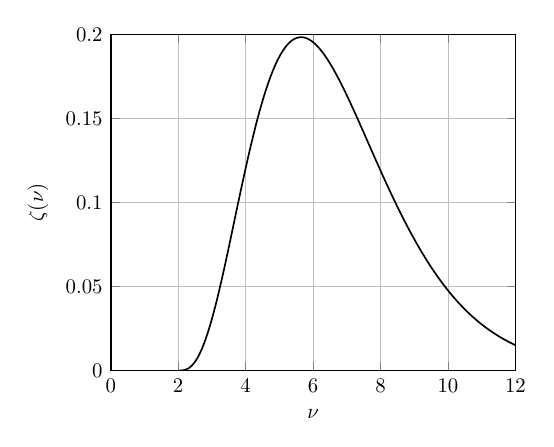}
\raisebox{-2.5pt}{\includegraphics[width=.49\textwidth]{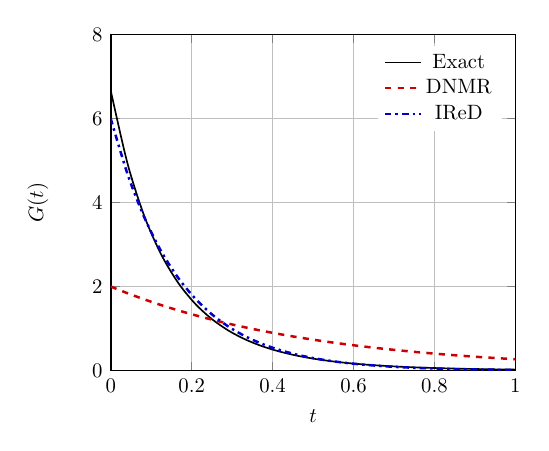}}
\caption{\label{figuz99} Left panel: A simple example of a spectral density of susceptibility, as given by equation \eqref{SPECTRAL}. Right panel: The associated exact Green function (solid line) and its Israel-Stewart approximations (dashed and dashdotted lines).}
\end{figure}
\begin{equation}\label{SPECTRAL}
    \zeta(\nu) =\dfrac{\nu}{36} \,  (\nu -2)^3 e^{-(\nu-2)}\Theta(\nu -2) \, ,
\end{equation}
whose profile is plotted in Fig. \ref{figuz99} (left panel). The Green function associated with this spectral density is (for $t{>}0$)
\begin{equation}
    G(t) = \dfrac{2(t^2+6t+10) \, e^{-2t}}{3(t+1)^6} \, ,
\end{equation}
which differs from an exponential by a rational factor. Nevertheless, our main results still hold. For example, IReD still provides a good approximation for the exact Green function, while DNMR is considerably less accurate (see figure \ref{figuz99}). The exact values of the Israel-Stewart transport coefficients for this model are reported below:
\begin{equation}
    \begin{split}
\zeta ={}& \int_0^{\infty} \zeta(\nu) \d\nu = 1 \, , \\
\tau_{\text{IR}}={}& \dfrac{1}{\zeta} \int_{0}^{\infty} \zeta(\nu) \dfrac{\d\nu}{\nu} = \dfrac{1}{6} \, ,\\
\tau_{\text{D}}={}& \max\left\{\frac{1}{\nu}\right\} = \dfrac{1}{2} \, . \\
    \end{split}
\end{equation}
The IReD Green function is just $6e^{-6t}$, while the DNMR Green function is $2e^{-2t}$. The analogies between figure \ref{figuz99} and figure \ref{figuz1} are quite remarkable. This confirms that, for the purposes of the present analysis, there is no fundamental difference between a simple model with three eigenvalues and a system with a continuous spectrum. 

% \subsection{Historical criticism of interpretations (ii) and (iii)}

% Geroch's claim that all ``Israel-Stewart phenomenology'' is unobservable has received considerable support over the decades. Part of this support goes back to some additional arguments, which may point to some intrinsic limitations of interpretations (ii) and (iii). Let us discuss these arguments in detail, and see if they modify our main results.

\subsection{Effect of stochastic fluctuations}

Deterministic descriptions like the Boltzmann equation and our model \eqref{SuperModel} neglect stochastic fluctuations. One would expect that, as long as the number of particles in a fluid element is very large, the impact of such fluctuations should be negligible. However, this is not always true. Studying the backreaction of stochastic sound waves on the flow over a very long time (i.e. in the limit $\omega \rightarrow 0$), one finds a cumulative effect that is larger than the second-order gradient corrections \cite{KovtunStickiness2011}. In our simple linear response model, one may picture this correction as an additional non-analytic power $\omega^{1/2}$ in equation \eqref{bornone}:
\begin{equation}
    \dfrac{\Pi(\omega)}{\theta(\omega)} \overset{\text{exact}}{=} -\gamma \kappa+\beta \omega^{1/2} -i\gamma \tau \kappa \, \omega +...
\end{equation}
This implies that, strictly speaking, interpretation (ii) is inconsistent with statistical mechanics. In fact, in the rigorous low-$\omega$ regime, all ``second-order phenomenology'' is covered by the effect of fluctuations, and is therefore undetectable. In practice, the actual relevance of the term $\beta \omega^{1/2}$ for physics is more limited than it may seem, since all real experiments have a finite duration, meaning that the rigorous low-$\omega$ regime is never achieved. Nevertheless, this effect is there, and it may become important in QCD \cite{KovtunStickiness2011}.

Although our analysis does not include the effect of fluctuations, the main conclusion that second-order physics is undetectable is strongly corroborated by our analysis. In fact, also in our non-stochastic model, second-order hydrodynamics is accurate only when Navier-Stokes is already quite accurate, making second-order corrections tiny, and hard to isolate. Hence, we confirm that interpretation (ii) {often gives} way to interpretation (i). 
Of course, this discussion has no impact on interpretation (iii), since transient hydrodynamics is concerned with the behavior of the system for short times ($\omega^{-1} \approx \tau_\Pi$), and it makes no claim of being more accurate than Navier-Stokes at late times.

%\subsubsection{Israel-Stewart phenomena are microscopic}

% The second criticism is well summarised in the following quote by \citet{Geroch2001}:

% ``\textit{let the Navier-Stokes fluid be nitrogen gas, and consider
% a sample of this fluid, at approximately standard temperature and pressure, for which changes in fluid state take place only over distances of the order of centimeters. For this sample, we know, more or less, what fluid velocity and temperature mean, i.e., we could construct centimeter-sized probes to measure them. But now suppose that, instead, our sample manifested state-changes taking place on distance scales of the order of $10^{-6}$ cm, the mean
% free path of the nitrogen molecules. In this case, things are not so simple: We must construct probes that work on these smaller scales. What typically happens is that different types of probes (...) will begin to give different results on the scale of $10^{-6}$ cm.}''

\section{Conclusions}

Our analysis suggests that we {may need to} revise some of the current paradigms of relativistic hydrodynamics.

First of all, we found that interpretation (i) {usually} prevails over interpretation (ii). In fact, second-order hydrodynamics (namely, Israel-Stewart viewed as a gradient expansion \cite{Baier2008}) is applicable only when Navier-Stokes is already pretty accurate, see figure \ref{fig:plots_d}, right panel. As soon as we exit the regime of applicability of first-order hydrodynamics (the ``asymptotic regime''), we also exit the regime of applicability of second-order hydrodynamics, which {often} turns out to behave even worse than Navier-Stokes, see figure \ref{fig:plots_kin}, right panel. This makes second-order gradient corrections extremely difficult to isolate phenomenologically, thereby corroborating the claims of Geroch \cite{Geroch2001} that second-order phenomena may be effectively undetectable in practice.

By contrast, interpretation (iii) was proven to be a perfectly valid approach, at least as an approximation of kinetic theory. In fact, the DNMR interpretation of Israel-Stewart hydrodynamics \cite{Denicol2012Boltzmann} fully passed ``Geroch's test'', being visibly more accurate than Navier-Stokes also in regimes where Navier-Stokes is manifestly non-reliable, see e.g. figure \ref{fig:plots_kin}, left panel. This implies that we can define for transient hydrodynamics a clear-cut regime of applicability, which extends far beyond that of first-order hydrodynamics. In other words, transient hydrodynamics is falsifiable in a rigorous sense, and it is a well-posed scientific theory. Indeed, in their seminal paper \cite{Israel_Stewart_1979}, Israel and Stewart themselves were viewing their own theory from the perspective of transient hydrodynamics, although they employed the 14-moment approximation to compute the transport coefficients, thereby obtaining inaccurate predictions for the transport coefficients, see figure \ref{fig:plots_kin} (``1AA'' curve). 

Unfortunately, we also found that neglecting the $\mathcal{K}$-terms from DNMR can considerably reduce its overall accuracy, in agreement with Ref. \cite{Denicol:2012vq}. This phenomenon can be dramatic for certain ``anomalous'' collision kernels, such as that of figure \ref{fig:plots_300} (see curve ``tDNMR''), where DNMR becomes even less reliable than Navier-Stokes in the transient regime. However, in commonplace situations such as figure \ref{fig:plots_kin}, DNMR still performs well enough also without the $\mathcal{K}$-terms.

Yet, the most important (and surprising) finding of the present article is probably the discovery that interpretations (ii) and (iii) can be unified within the IReD theory \cite{WagnerIReD2022}. In fact, IReD fulfills interpretation (ii) by construction, see figure \ref{fig:plots_d}. Furthermore, due to a fortuitous ``twist of fate'', its relaxation time happens to be the average of all the (possibly infinite) relaxation times of kinetic theory, weighted on their susceptibility to the hydrodynamic gradients, see equation \eqref{MagicIRED}. This weighted average constitutes an excellent fit of the microscopic Green function, making IReD reliable for any collision kernel (even the anomalous ones considered figures \ref{figuz1}, \ref{fig:plots_300}, and \ref{figuz99}). As a consequence, by keeping track of all relaxation times, IReD can reach levels of accuracy that escape DNMR. For example, in the realistic case shown in figure \ref{fig:plots_kin}, the error of IReD theory is an order of magnitude smaller than that of DNMR (even keeping the $\mathcal{K}$-terms), and it is two orders of magnitude smaller than that of Navier-Stokes (see table \ref{tbl:scores_kin}). For these reasons, we believe that switching from the DNMR to the IReD paradigm might significantly improve the accuracy of hydrodynamic simulations.

In summary, we hope we convinced the reader that Israel-Stewart hydrodynamics constitutes an objective improvement to Navier-Stokes, at least within the kinetic theory framework. Its regime of applicability extends to processes whose timescale is comparable to the relaxation time, provided that the transport coefficients are computed using the IReD prescription. In our opinion, this could help resolve a debate that has been going on since 1978 \cite{Israel_Stewart_1979}.

\section*{Acknowledgements}

This work was supported by a Vanderbilt's Seeding Success Grant.  This research was supported in part by the National Science Foundation under Grant No. PHY-1748958. 
D.W.\ acknowledges support by the Research Cluster ELEMENTS (Project ID 500/10.006), the Deutsche
Forschungsgemeinschaft (DFG, German Research Foundation) through the 
CRC-TR 211 ``Strong-interaction matter
under extreme conditions'' -- project number 315477589 -- TRR 211, and the 
Ministry of Research, Innovation and Digitization, CNCS - UEFISCDI,
project number PN-III-P1-1.1-TE-2021-1707, within PNCDI III. We thank J. Noronha and M. Stephanov for reading our manuscript and providing useful comments.

\appendix

\section{Connection to kinetic theory}
\label{app:kin_theory}
The purpose of this appendix is to clarify the connection of our simplified model \eqref{SuperModel} to kinetic theory, which is often the microscopic starting point for deriving hydrodynamics. For simplicity, we will consider a gas of classical particles.

The energy-momentum tensor $T^{\mu\nu}$ and the particle four-current $N^\mu$ are the basic quantity of interest in a fluid which admits a kinetic description as mentioned above.
Assuming Landau matching conditions, i.e., demanding that there be no energy flux in the fluid-rest frame, we can decompose $T^{\mu\nu}$ and $N^\mu$ with respect to the fluid four-velocity $u^\mu$ as
\begin{equation}
    T^{\mu\nu}=\varepsilon u^\mu u^\nu - \Delta^{\mu\nu} \left(P+\Pi\right) +\pi^{\mu\nu}\;,\qquad N^\mu = n u^\mu + n^\mu \;,
\end{equation}
with $\Delta^{\mu\nu}=g^{\mu\nu}-u^\mu u^\nu$.
Here, $\varepsilon$, $P$, and $n$ are the (equilibrium) energy density, pressure, and particle-number density, respectively, which determine ideal fluid dynamics. In contrast, additionally computing the bulk viscous pressure $\Pi$, the particle-diffusion current $n^\mu$, and the shear-stress tensor $\pi^{\mu\nu}$ is the objective of dissipative hydrodynamics. 

These dissipative quantities admit the following kinetic description \cite{Denicol2012Boltzmann},
\begin{equation}
    \Pi=-\frac{m^2}{3} \int \d K \delta f_\k\;,\qquad n^\mu = \int \d K k^{\langle\mu\rangle}\delta f_\k\;,\qquad\pi^{\mu\nu}= \int \d K \, k^{\langle\mu} k^{\nu\rangle} \delta f_\k \;,
\end{equation}
where the angular brackets symbolize that the respective tensor is projected to be orthogonal to the four-velocity $u^\mu$ as well as symmetric and traceless in all pairs of indices \cite{Denicol2012Boltzmann}.
Furthermore, $\d K = \d^3 \k /[(2\pi)^3 k^0]$ is the measure in momentum space, $k^0=\sqrt{\k^2+m^2}$ the on-shell energy, $E_\k=u^\mu k_\mu$ the energy in the rest frame of the fluid, and $\delta f_\k$ denotes the deviation from the local-equilibrium single-particle distribution function $f_{0\k}=e^{\alpha-\beta E_\k}$, where $\beta=1/T$ is the inverse temperature and  $\alpha=\beta \mu$ denotes the ratio of chemical potential over temperature.

While the evolution of this function is determined by the Boltzmann equation, when deriving dissipative hydrodynamics it is advantageous to instead track the so-called irreducible moments of $\delta f_\k$,
\begin{equation}
    \rho_r^{\mu_1\cdots \mu_\ell}= \int \d K  E_\k^{r} k^{\langle\mu_1}\cdots k^{\mu_\ell\rangle} \delta f_\k\;.
\end{equation}
Using these moments, we can identify $\Pi=-(m^2/3)\rho_0$, $n^\mu=\rho_0^\mu$, and $\pi^{\mu\nu}=\rho_0^{\mu\nu}$.
Denoting the comoving derivative of a quantity $A$ with a dot, $\dot{A}=u^\mu \partial_\mu A$, the equations of motion for the irreducible moments of tensor-ranks $\ell\leq 2$ read \cite{Denicol2012Boltzmann}
\begin{subequations}\label{eq:rhodot_all}
\begin{align}
 \dot{\rho}_r - C_{r-1} =&\, \alpha_r^{(0)} \theta -
 \frac{G_{2r}}{D_{20}} \Pi \theta + 
 \frac{G_{2r}}{D_{20}} \pi^{\mu\nu} \sigma_{\mu\nu} + 
 \frac{G_{3r}}{D_{20}} \partial_\mu n^\mu
 + (r - 1) \rho^{\mu\nu}_{r-2} \sigma_{\mu\nu} + 
 r \rho^\mu_{r-1} \dot{u}_\mu - \nabla_\mu \rho^\mu_{r-1}\nonumber\\
 &- \frac{1}{3}\left[(r+2) \rho_r - (r -1)m^2 \rho_{r-2}\right]
 \theta\;, \label{eq:rhodot_l0}\\
 \dot{\rho}^{\langle\mu\rangle}_r - C_{r-1}^{\langle \mu \rangle} =&\, \alpha_r^{(1)} I^\mu + 
 \rho^\nu_r \omega^\mu{}_\nu + 
 \frac{1}{3}[(r-1) m^2 \rho^\mu_{r-2} - (r+ 3) \rho^\mu_r] \theta 
 - \Delta^\mu_\lambda \nabla_\nu \rho^{\lambda \nu}_{r-1} 
 + r \rho^{\mu\nu}_{r-1} \dot{u}_\nu \nonumber\\
 &+ \frac{1}{5} \left[(2r-2) m^2 \rho^\nu_{r-2} - (2r+3) \rho^\nu_r\right] \sigma^\mu_\nu 
 + \frac{1}{3}\left[m^2 r \rho_{r-1} - (r+3) \rho_{r+1}\right] \dot{u}^\mu \nonumber\\
 &+ \frac{\beta I_{r+2,1}}{\varepsilon + P} 
 (\Pi \dot{u}^\mu - \nabla^\mu \Pi + \Delta^\mu_\nu \partial_\lambda \pi^{\lambda \nu}) 
 -\frac{1}{3} \nabla^\mu (m^2 \rho_{r-1} - \rho_{r+1}) 
 + (r -1 ) \rho^{\mu\nu\lambda}_{r-2} \sigma_{\lambda\nu}\;,
 \label{eq:rhodot_l1} \\
 \dot{\rho}^{\langle \mu\nu\rangle}_r - C^{\langle \mu\nu \rangle}_{r-1} =& \,
 2 \alpha^{(2)}_r \sigma^{\mu\nu} - \frac{2}{7} \left[(2r+5) \rho^{\lambda\langle \mu}_r - 2m^2(r-1) \rho^{\lambda\langle \mu}_{r-2}\right] \sigma^{\nu\rangle}_\lambda + 
 2\rho^{\lambda\langle \mu}_r \omega^{\nu\rangle}{}_\lambda \nonumber\\
 &+ \frac{2}{15}[(r+4)\rho_{r+2} - (2r+3) m^2 \rho_r + 
 (r - 1)m^4 \rho_{r-2}] \sigma^{\mu\nu} + 
 \frac{2}{5} \nabla^{\langle \mu} (\rho^{\nu \rangle}_{r+1} - m^2 \rho^{\nu\rangle}_{r-1}) \nonumber\\
 &- \frac{2}{5} \left[(r+5) \rho^{\langle \mu}_{r+1} - r m^2 \rho^{\langle \mu}_{r-1}\right] \dot{u}^{\nu \rangle} 
 - \frac{1}{3} \left[(r + 4) \rho^{\mu\nu}_r - m^2 (r-1) \rho^{\mu\nu}_{r-2}\right] \theta 
 \nonumber\\
 & + (r - 1) \rho^{\mu\nu\lambda\rho}_{r-2} \sigma_{\lambda \rho} - \Delta^{\mu\nu}_{\alpha\beta} 
 \nabla_\lambda \rho^{\alpha\beta\lambda}_{r-1} + 
 r \rho^{\mu\nu\lambda}_{r-1} \dot{u}_\lambda\;,
 \label{eq:rhodot_l2}
\end{align}
\end{subequations}
where $I^\mu=\nabla^\mu \alpha$ (with the spacelike gradient $\nabla^\mu = \Delta^{\mu\nu}\partial_\nu$), $\sigma_{\mu\nu}=\nabla_{\langle\mu} u_{\nu\rangle}$ and $\omega_{\mu\nu}=\frac{1}{2} (\partial_\mu u_\nu-\partial_\nu u_\mu)$ are the shear and vorticity tensors, respectively, and $\theta=\nabla_\mu u^\mu$ denotes the expansion scalar. The quantity
\begin{equation}
 C^{\langle \mu_1 \cdots \mu_\ell \rangle}_r = 
 \int \d K \, E_\k^r k^{\langle \mu_1} \cdots k^{\mu_\ell \rangle} 
 C[f]\;.
 \label{eq:Cr_def}
\end{equation}
is an irreducible moment of the collision term. Moreover, we have introduced the basic thermodynamic integral
\begin{equation}
    I_{nq} = \frac{1}{(2q+1)!!} \int \d K\, E_\k^{n-2q} (-\Delta^{\alpha\beta} k_\alpha k_\beta)^q f_{0\k}\;,\label{eq:I_def}
\end{equation}
and the combinations $G_{nm} = I_{n0} I_{m0} - I_{n-1,0} I_{m+1,0}$, $D_{nq} = I_{n+1,q} I_{n-1,q} - I_{nq}^2$. 
The coefficients appearing in the first terms on the right-hand sides of equations \eqref{eq:rhodot_all} are given as
\begin{equation}
 \alpha_r^{(0)} = (1 - r) I_{r1} - I_{r0} - \frac{1}{D_{20}} 
 [G_{2r}(\varepsilon + P) - G_{3r} n]\;,\qquad
 \alpha_r^{(1)} = I_{r+1,1} - \frac{n}{\varepsilon + P} I_{r+2,1}\;,\qquad
 \alpha_r^{(2)} = I_{r+2,1} + (r - 1) I_{r+2,2}\;.
\end{equation}
When neglecting the nonlinear contributions in equations \eqref{eq:rhodot_all}, we find
\begin{subequations}\label{eq:rhodot_all_linear}
\begin{align}
 \dot{\rho}_r +\sum_n \mathcal{A}_{rn}^{(0)} \rho_n =&\, \alpha_r^{(0)} \theta  + 
 \frac{G_{3r}}{D_{20}} \nabla_\mu n^\mu
 - \nabla_\mu \rho^\mu_{r-1}\label{eq:rhodot_l0_linear}\\
 \dot{\rho}^{\langle\mu\rangle}_r+ \sum_n \mathcal{A}_{rn}^{(1)} \rho_n^\mu =&\, \alpha_r^{(1)} I^\mu 
 - \Delta^\mu_\lambda \nabla_\nu \rho^{\lambda \nu}_{r-1} 
 + \frac{\beta I_{r+2,1}}{\varepsilon + P} 
 (\Delta^\mu_\nu \nabla_\lambda \pi^{\lambda \nu}- \nabla^\mu \Pi ) 
 -\frac{1}{3} \nabla^\mu (m^2 \rho_{r-1} - \rho_{r+1}) 
 \;,
 \label{eq:rhodot_l1_linear} \\
 \dot{\rho}^{\langle \mu\nu\rangle}_r + \sum_n \mathcal{A}_{rn}^{(2)} \rho_n^{\mu\nu} =& \,
 2 \alpha^{(2)}_r \sigma^{\mu\nu}+ 
 \frac{2}{5} \nabla^{\langle \mu} (\rho^{\nu \rangle}_{r+1} - m^2 \rho^{\nu\rangle}_{r-1})  - \Delta^{\mu\nu}_{\alpha\beta} 
 \nabla_\lambda \rho^{\alpha\beta\lambda}_{r-1} \;,
 \label{eq:rhodot_l2_linear}
\end{align}
\end{subequations}
where we employed that the irreducible moments of the collision term can be written as
\begin{equation}
    C_{r-1}^{\langle\mu_1\cdots \mu_\ell\rangle}= -\sum_{n}\mathcal{A}_{rn}^{(\ell)}\rho_n^{\mu_1\cdots\mu_\ell}+\text{nonlinear terms}\;.
\end{equation}
Inverting the linearized collision matrices, $\tau^{(\ell)}=(\mathcal{A}^{(\ell)})^{-1}$, we then obtain
\begin{subequations}\label{eq:rhodot_all_linear_2}
\begin{align}
 \sum_n \tau_{rn}^{(0)}\dot{\rho}_n + \rho_r =&\,\sum_n \tau_{rn}^{(0)}\left( \alpha_n^{(0)} \theta  + 
 \frac{G_{3r}}{D_{20}} \nabla_\mu n^\mu
 - \nabla_\mu \rho^\mu_{n-1}\right)\label{eq:rhodot_l0_linear_2}\\
 \sum_n \tau_{rn}^{(1)}\dot{\rho}^{\langle\mu\rangle}_n+  \rho_r^\mu =&\,\sum_n \tau_{rn}^{(1)}\left[ \alpha_n^{(1)} I^\mu 
 - \Delta^\mu_\lambda \nabla_\nu \rho^{\lambda \nu}_{n-1} 
 + \frac{\beta I_{n+2,1}}{\varepsilon + P} 
 (\Delta^\mu_\nu \nabla_\lambda \pi^{\lambda \nu}- \nabla^\mu \Pi ) 
 -\frac{1}{3} \nabla^\mu (m^2 \rho_{n-1} - \rho_{n+1}) \right]
 \;,
 \label{eq:rhodot_l1_linear_2} \\
 \sum_n \tau_{rn}^{(2)} \dot{\rho}^{\langle \mu\nu\rangle}_n + \rho_r^{\mu\nu} =& \, \sum_n \tau_{rn}^{(2)}\left[
 2 \alpha^{(2)}_n \sigma^{\mu\nu}+ 
 \frac{2}{5} \nabla^{\langle \mu} (\rho^{\nu \rangle}_{n+1} - m^2 \rho^{\nu\rangle}_{n-1})  - \Delta^{\mu\nu}_{\alpha\beta} 
 \nabla_\lambda \rho^{\alpha\beta\lambda}_{n-1}\right] \;,
 \label{eq:rhodot_l2_linear_2}
\end{align}
\end{subequations}
At this point, we can make a connection between the linearized moment equations \eqref{eq:rhodot_all_linear_2} and our basic model \eqref{SuperModel}. Upon evaluating the moment equations along the worldline of a specific fluid element, as schematically depicted in figure \ref{fig:MDL}, the moments will only depend on the proper time $t$ that parametrizes the wordline. Then, we can identify the quantities $\mathbb{A}$ in equation \eqref{SuperModel} with the (in principle infinite-dimensional) vectors of irreducible moments,\footnote{The moments $\rho_1$, $\rho_2$, and $\rho_1^\mu$ do not appear due to Landau matching.} i.e., $\mathbb{A}=\{\rho_0,\rho_3,\ldots\}^\mathrm{T}$ for $\ell=0$, $\mathbb{A}^\mu=\{\rho_0^\mu,\rho_2^\mu,\ldots\}^\mathrm{T}$ for $\ell=1$, and  $\mathbb{A}^{\mu\nu}=\{\rho_0^{\mu\nu},\rho_1^{\mu\nu},\ldots\}^\mathrm{T}$ for $\ell=2$. Similarly, the matrices $\tau$ can be identified with the respective inverse linearized collision matrices, $\tau=\tau^{(\ell)}$. The vectors $\gamma$ that specify which combination of the moments is of interest for the conserved currents are given by $\gamma=\{m^2/3,0,\ldots\}^\mathrm{T}$ for $\ell=0$, and $\gamma=\{-1,0,\ldots\}^\mathrm{T}$ for $\ell=1$ or $\ell=2$.\\
The set of spatial gradients on the right-hand sides denote the external driving forces that influence the proper-time evolution of the quantities $\mathbb{A}$. While, as is evident from equations \eqref{eq:rhodot_all_linear_2}, there can be a multitude of these terms, one simplification of the model \eqref{SuperModel} lies in collecting all of them in one external driving force $\theta(t)$. Note that in neglecting nonlinear terms we also assume spatial gradients to be small, thereby excluding regions of figure \ref{fig:applicabluz} that are far away from the vertical axis.
Upon considering a fluid of massless uncharged particles and neglecting the second and third term on the right-hand side of Eq. \eqref{eq:rhodot_l2_linear_2}, we arrive at the setup considered in section \ref{sec:kin}.

\bibliography{Biblio}

\begin{thebibliography}{43}%
\makeatletter
\providecommand \@ifxundefined [1]{%
 \@ifx{#1\undefined}
}%
\providecommand \@ifnum [1]{%
 \ifnum #1\expandafter \@firstoftwo
 \else \expandafter \@secondoftwo
 \fi
}%
\providecommand \@ifx [1]{%
 \ifx #1\expandafter \@firstoftwo
 \else \expandafter \@secondoftwo
 \fi
}%
\providecommand \natexlab [1]{#1}%
\providecommand \enquote  [1]{``#1''}%
\providecommand \bibnamefont  [1]{#1}%
\providecommand \bibfnamefont [1]{#1}%
\providecommand \citenamefont [1]{#1}%
\providecommand \href@noop [0]{\@secondoftwo}%
\providecommand \href [0]{\begingroup \@sanitize@url \@href}%
\providecommand \@href[1]{\@@startlink{#1}\@@href}%
\providecommand \@@href[1]{\endgroup#1\@@endlink}%
\providecommand \@sanitize@url [0]{\catcode `\\12\catcode `\$12\catcode
  `\&12\catcode `\#12\catcode `\^12\catcode `\_12\catcode `\%12\relax}%
\providecommand \@@startlink[1]{}%
\providecommand \@@endlink[0]{}%
\providecommand \url  [0]{\begingroup\@sanitize@url \@url }%
\providecommand \@url [1]{\endgroup\@href {#1}{\urlprefix }}%
\providecommand \urlprefix  [0]{URL }%
\providecommand \Eprint [0]{\href }%
\providecommand \doibase [0]{http://dx.doi.org/}%
\providecommand \selectlanguage [0]{\@gobble}%
\providecommand \bibinfo  [0]{\@secondoftwo}%
\providecommand \bibfield  [0]{\@secondoftwo}%
\providecommand \translation [1]{[#1]}%
\providecommand \BibitemOpen [0]{}%
\providecommand \bibitemStop [0]{}%
\providecommand \bibitemNoStop [0]{.\EOS\space}%
\providecommand \EOS [0]{\spacefactor3000\relax}%
\providecommand \BibitemShut  [1]{\csname bibitem#1\endcsname}%
\let\auto@bib@innerbib\@empty
%</preamble>
\bibitem [{\citenamefont {Israel}\ and\ \citenamefont
  {Stewart}(1979)}]{Israel_Stewart_1979}%
  \BibitemOpen
  \bibfield  {author} {\bibinfo {author} {\bibfnamefont {W.}~\bibnamefont
  {Israel}}\ and\ \bibinfo {author} {\bibfnamefont {J.}~\bibnamefont
  {Stewart}},\ }\href {\doibase https://doi.org/10.1016/0003-4916(79)90130-1}
  {\bibfield  {journal} {\bibinfo  {journal} {Annals of Physics}\ }\textbf
  {\bibinfo {volume} {118}},\ \bibinfo {pages} {341 } (\bibinfo {year}
  {1979})}\BibitemShut {NoStop}%
\bibitem [{\citenamefont {Hiscock}\ and\ \citenamefont
  {Lindblom}(1983)}]{Hishcock1983}%
  \BibitemOpen
  \bibfield  {author} {\bibinfo {author} {\bibfnamefont {W.~A.}\ \bibnamefont
  {Hiscock}}\ and\ \bibinfo {author} {\bibfnamefont {L.}~\bibnamefont
  {Lindblom}},\ }\href {\doibase https://doi.org/10.1016/0003-4916(83)90288-9}
  {\bibfield  {journal} {\bibinfo  {journal} {Annals of Physics}\ }\textbf
  {\bibinfo {volume} {151}},\ \bibinfo {pages} {466 } (\bibinfo {year}
  {1983})}\BibitemShut {NoStop}%
\bibitem [{\citenamefont {{Olson}}(1990)}]{OlsonLifsh1990}%
  \BibitemOpen
  \bibfield  {author} {\bibinfo {author} {\bibfnamefont {T.~S.}\ \bibnamefont
  {{Olson}}},\ }\href {\doibase 10.1016/0003-4916(90)90366-V} {\bibfield
  {journal} {\bibinfo  {journal} {Annals of Physics}\ }\textbf {\bibinfo
  {volume} {199}},\ \bibinfo {pages} {18} (\bibinfo {year} {1990})}\BibitemShut
  {NoStop}%
\bibitem [{\citenamefont {{Baier}}\ \emph {et~al.}(2008)\citenamefont
  {{Baier}}, \citenamefont {{Romatschke}}, \citenamefont {{Thanh Son}},
  \citenamefont {{Starinets}},\ and\ \citenamefont {{Stephanov}}}]{Baier2008}%
  \BibitemOpen
  \bibfield  {author} {\bibinfo {author} {\bibfnamefont {R.}~\bibnamefont
  {{Baier}}}, \bibinfo {author} {\bibfnamefont {P.}~\bibnamefont
  {{Romatschke}}}, \bibinfo {author} {\bibfnamefont {D.}~\bibnamefont {{Thanh
  Son}}}, \bibinfo {author} {\bibfnamefont {A.~O.}\ \bibnamefont
  {{Starinets}}}, \ and\ \bibinfo {author} {\bibfnamefont {M.~A.}\ \bibnamefont
  {{Stephanov}}},\ }\href {\doibase 10.1088/1126-6708/2008/04/100} {\bibfield
  {journal} {\bibinfo  {journal} {Journal of High Energy Physics}\ }\textbf
  {\bibinfo {volume} {2008}},\ \bibinfo {eid} {100} (\bibinfo {year} {2008})},\
  \Eprint {http://arxiv.org/abs/0712.2451} {arXiv:0712.2451 [hep-th]}
  \BibitemShut {NoStop}%
\bibitem [{\citenamefont {Denicol}\ \emph {et~al.}(2012)\citenamefont
  {Denicol}, \citenamefont {Niemi}, \citenamefont {Moln\'ar},\ and\
  \citenamefont {Rischke}}]{Denicol2012Boltzmann}%
  \BibitemOpen
  \bibfield  {author} {\bibinfo {author} {\bibfnamefont {G.~S.}\ \bibnamefont
  {Denicol}}, \bibinfo {author} {\bibfnamefont {H.}~\bibnamefont {Niemi}},
  \bibinfo {author} {\bibfnamefont {E.}~\bibnamefont {Moln\'ar}}, \ and\
  \bibinfo {author} {\bibfnamefont {D.~H.}\ \bibnamefont {Rischke}},\ }\href
  {\doibase 10.1103/PhysRevD.85.114047} {\bibfield  {journal} {\bibinfo
  {journal} {Phys. Rev. D}\ }\textbf {\bibinfo {volume} {85}},\ \bibinfo
  {pages} {114047} (\bibinfo {year} {2012})}\BibitemShut {NoStop}%
\bibitem [{\citenamefont {{Gavassino}}\ and\ \citenamefont
  {{Antonelli}}(2023)}]{GavassinoGENERIC2022}%
  \BibitemOpen
  \bibfield  {author} {\bibinfo {author} {\bibfnamefont {L.}~\bibnamefont
  {{Gavassino}}}\ and\ \bibinfo {author} {\bibfnamefont {M.}~\bibnamefont
  {{Antonelli}}},\ }\href {\doibase 10.1088/1361-6382/acc165} {\bibfield
  {journal} {\bibinfo  {journal} {Classical and Quantum Gravity}\ }\textbf
  {\bibinfo {volume} {40}},\ \bibinfo {eid} {075012} (\bibinfo {year}
  {2023})},\ \Eprint {http://arxiv.org/abs/2209.12865} {arXiv:2209.12865
  [gr-qc]} \BibitemShut {NoStop}%
\bibitem [{\citenamefont {{Geroch}}(1995)}]{Geroch1995}%
  \BibitemOpen
  \bibfield  {author} {\bibinfo {author} {\bibfnamefont {R.}~\bibnamefont
  {{Geroch}}},\ }\href {\doibase 10.1063/1.530958} {\bibfield  {journal}
  {\bibinfo  {journal} {Journal of Mathematical Physics}\ }\textbf {\bibinfo
  {volume} {36}},\ \bibinfo {pages} {4226} (\bibinfo {year}
  {1995})}\BibitemShut {NoStop}%
\bibitem [{\citenamefont {{Lindblom}}(1996)}]{LindblomRelaxation1996}%
  \BibitemOpen
  \bibfield  {author} {\bibinfo {author} {\bibfnamefont {L.}~\bibnamefont
  {{Lindblom}}},\ }\href {\doibase 10.1006/aphy.1996.0036} {\bibfield
  {journal} {\bibinfo  {journal} {Annals of Physics}\ }\textbf {\bibinfo
  {volume} {247}},\ \bibinfo {pages} {1} (\bibinfo {year} {1996})},\ \Eprint
  {http://arxiv.org/abs/gr-qc/9508058} {arXiv:gr-qc/9508058 [gr-qc]}
  \BibitemShut {NoStop}%
\bibitem [{\citenamefont {{Geroch}}(2001)}]{Geroch2001}%
  \BibitemOpen
  \bibfield  {author} {\bibinfo {author} {\bibfnamefont {R.}~\bibnamefont
  {{Geroch}}},\ }\href {\doibase 10.48550/arXiv.gr-qc/0103112} {\bibfield
  {journal} {\bibinfo  {journal} {arXiv e-prints}\ ,\ \bibinfo {eid}
  {gr-qc/0103112}} (\bibinfo {year} {2001})},\ \Eprint
  {http://arxiv.org/abs/gr-qc/0103112} {arXiv:gr-qc/0103112 [gr-qc]}
  \BibitemShut {NoStop}%
\bibitem [{\citenamefont {{Kost{\"a}dt}}\ and\ \citenamefont
  {{Liu}}(2000)}]{Kost2000}%
  \BibitemOpen
  \bibfield  {author} {\bibinfo {author} {\bibfnamefont {P.}~\bibnamefont
  {{Kost{\"a}dt}}}\ and\ \bibinfo {author} {\bibfnamefont {M.}~\bibnamefont
  {{Liu}}},\ }\href {\doibase 10.1103/PhysRevD.62.023003} {\bibfield  {journal}
  {\bibinfo  {journal} {\prd}\ }\textbf {\bibinfo {volume} {62}},\ \bibinfo
  {eid} {023003} (\bibinfo {year} {2000})},\ \Eprint
  {http://arxiv.org/abs/cond-mat/0010276} {arXiv:cond-mat/0010276
  [cond-mat.stat-mech]} \BibitemShut {NoStop}%
\bibitem [{\citenamefont {{Romatschke}}\ and\ \citenamefont
  {{Romatschke}}(2017)}]{Romatschke2017}%
  \BibitemOpen
  \bibfield  {author} {\bibinfo {author} {\bibfnamefont {P.}~\bibnamefont
  {{Romatschke}}}\ and\ \bibinfo {author} {\bibfnamefont {U.}~\bibnamefont
  {{Romatschke}}},\ }\href {\doibase 10.48550/arXiv.1712.05815} {\bibfield
  {journal} {\bibinfo  {journal} {arXiv e-prints}\ ,\ \bibinfo {eid}
  {arXiv:1712.05815}} (\bibinfo {year} {2017})},\ \Eprint
  {http://arxiv.org/abs/1712.05815} {arXiv:1712.05815 [nucl-th]} \BibitemShut
  {NoStop}%
\bibitem [{\citenamefont {{Florkowski}}\ \emph {et~al.}(2018)\citenamefont
  {{Florkowski}}, \citenamefont {{Heller}},\ and\ \citenamefont
  {{Spali{\'n}ski}}}]{FlorkowskiReview2018}%
  \BibitemOpen
  \bibfield  {author} {\bibinfo {author} {\bibfnamefont {W.}~\bibnamefont
  {{Florkowski}}}, \bibinfo {author} {\bibfnamefont {M.~P.}\ \bibnamefont
  {{Heller}}}, \ and\ \bibinfo {author} {\bibfnamefont {M.}~\bibnamefont
  {{Spali{\'n}ski}}},\ }\href {\doibase 10.1088/1361-6633/aaa091} {\bibfield
  {journal} {\bibinfo  {journal} {Reports on Progress in Physics}\ }\textbf
  {\bibinfo {volume} {81}},\ \bibinfo {eid} {046001} (\bibinfo {year}
  {2018})},\ \Eprint {http://arxiv.org/abs/1707.02282} {arXiv:1707.02282
  [hep-ph]} \BibitemShut {NoStop}%
\bibitem [{\citenamefont {Denicol}\ \emph {et~al.}(2011)\citenamefont
  {Denicol}, \citenamefont {Noronha}, \citenamefont {Niemi},\ and\
  \citenamefont {Rischke}}]{Denicol_Relaxation_2011}%
  \BibitemOpen
  \bibfield  {author} {\bibinfo {author} {\bibfnamefont {G.~S.}\ \bibnamefont
  {Denicol}}, \bibinfo {author} {\bibfnamefont {J.}~\bibnamefont {Noronha}},
  \bibinfo {author} {\bibfnamefont {H.}~\bibnamefont {Niemi}}, \ and\ \bibinfo
  {author} {\bibfnamefont {D.~H.}\ \bibnamefont {Rischke}},\ }\href {\doibase
  10.1103/PhysRevD.83.074019} {\bibfield  {journal} {\bibinfo  {journal} {Phys.
  Rev. D}\ }\textbf {\bibinfo {volume} {83}},\ \bibinfo {pages} {074019}
  (\bibinfo {year} {2011})}\BibitemShut {NoStop}%
\bibitem [{\citenamefont {{Grozdanov}}\ and\ \citenamefont
  {{Polonyi}}(2013)}]{Grozdanov2013}%
  \BibitemOpen
  \bibfield  {author} {\bibinfo {author} {\bibfnamefont {S.}~\bibnamefont
  {{Grozdanov}}}\ and\ \bibinfo {author} {\bibfnamefont {J.}~\bibnamefont
  {{Polonyi}}},\ }\href@noop {} {\bibfield  {journal} {\bibinfo  {journal}
  {arXiv e-prints}\ ,\ \bibinfo {eid} {arXiv:1305.3670}} (\bibinfo {year}
  {2013})},\ \Eprint {http://arxiv.org/abs/1305.3670} {arXiv:1305.3670
  [hep-th]} \BibitemShut {NoStop}%
\bibitem [{\citenamefont {Gavassino}\ \emph {et~al.}(2022)\citenamefont
  {Gavassino}, \citenamefont {Antonelli},\ and\ \citenamefont
  {Haskell}}]{GavassinoNonHydro2022}%
  \BibitemOpen
  \bibfield  {author} {\bibinfo {author} {\bibfnamefont {L.}~\bibnamefont
  {Gavassino}}, \bibinfo {author} {\bibfnamefont {M.}~\bibnamefont
  {Antonelli}}, \ and\ \bibinfo {author} {\bibfnamefont {B.}~\bibnamefont
  {Haskell}},\ }\href {\doibase 10.1103/PhysRevD.106.056010} {\bibfield
  {journal} {\bibinfo  {journal} {Phys. Rev. D}\ }\textbf {\bibinfo {volume}
  {106}},\ \bibinfo {pages} {056010} (\bibinfo {year} {2022})}\BibitemShut
  {NoStop}%
\bibitem [{\citenamefont {Gavassino}\ and\ \citenamefont
  {Antonelli}(2021)}]{GavassinoFronntiers2021}%
  \BibitemOpen
  \bibfield  {author} {\bibinfo {author} {\bibfnamefont {L.}~\bibnamefont
  {Gavassino}}\ and\ \bibinfo {author} {\bibfnamefont {M.}~\bibnamefont
  {Antonelli}},\ }\href {\doibase 10.3389/fspas.2021.686344} {\bibfield
  {journal} {\bibinfo  {journal} {Front. Astron. Space Sci.}\ }\textbf
  {\bibinfo {volume} {8}},\ \bibinfo {pages} {686344} (\bibinfo {year}
  {2021})},\ \Eprint {http://arxiv.org/abs/2105.15184} {arXiv:2105.15184
  [gr-qc]} \BibitemShut {NoStop}%
\bibitem [{\citenamefont {Gavassino}(2022)}]{GavassinoSuperluminal2021}%
  \BibitemOpen
  \bibfield  {author} {\bibinfo {author} {\bibfnamefont {L.}~\bibnamefont
  {Gavassino}},\ }\href {\doibase 10.1103/PhysRevX.12.041001} {\bibfield
  {journal} {\bibinfo  {journal} {Phys. Rev. X}\ }\textbf {\bibinfo {volume}
  {12}},\ \bibinfo {pages} {041001} (\bibinfo {year} {2022})}\BibitemShut
  {NoStop}%
\bibitem [{\citenamefont {Gavassino}(2023)}]{GavassinoBounds2023}%
  \BibitemOpen
  \bibfield  {author} {\bibinfo {author} {\bibfnamefont {L.}~\bibnamefont
  {Gavassino}},\ }\href {\doibase 10.1016/j.physletb.2023.137854} {\bibfield
  {journal} {\bibinfo  {journal} {Phys. Lett. B}\ }\textbf {\bibinfo {volume}
  {840}},\ \bibinfo {pages} {137854} (\bibinfo {year} {2023})},\ \Eprint
  {http://arxiv.org/abs/2301.06651} {arXiv:2301.06651 [hep-th]} \BibitemShut
  {NoStop}%
\bibitem [{\citenamefont {Moln\'ar}\ \emph {et~al.}(2014)\citenamefont
  {Moln\'ar}, \citenamefont {Niemi}, \citenamefont {Denicol},\ and\
  \citenamefont {Rischke}}]{Molnar:2013lta}%
  \BibitemOpen
  \bibfield  {author} {\bibinfo {author} {\bibfnamefont {E.}~\bibnamefont
  {Moln\'ar}}, \bibinfo {author} {\bibfnamefont {H.}~\bibnamefont {Niemi}},
  \bibinfo {author} {\bibfnamefont {G.~S.}\ \bibnamefont {Denicol}}, \ and\
  \bibinfo {author} {\bibfnamefont {D.~H.}\ \bibnamefont {Rischke}},\ }\href
  {\doibase 10.1103/PhysRevD.89.074010} {\bibfield  {journal} {\bibinfo
  {journal} {Phys. Rev. D}\ }\textbf {\bibinfo {volume} {89}},\ \bibinfo
  {pages} {074010} (\bibinfo {year} {2014})},\ \Eprint
  {http://arxiv.org/abs/1308.0785} {arXiv:1308.0785 [nucl-th]} \BibitemShut
  {NoStop}%
\bibitem [{\citenamefont {Denicol}\ \emph
  {et~al.}(2014{\natexlab{a}})\citenamefont {Denicol}, \citenamefont {Niemi},
  \citenamefont {Bouras}, \citenamefont {Molnar}, \citenamefont {Xu},
  \citenamefont {Rischke},\ and\ \citenamefont {Greiner}}]{Denicol:2012vq}%
  \BibitemOpen
  \bibfield  {author} {\bibinfo {author} {\bibfnamefont {G.~S.}\ \bibnamefont
  {Denicol}}, \bibinfo {author} {\bibfnamefont {H.}~\bibnamefont {Niemi}},
  \bibinfo {author} {\bibfnamefont {I.}~\bibnamefont {Bouras}}, \bibinfo
  {author} {\bibfnamefont {E.}~\bibnamefont {Molnar}}, \bibinfo {author}
  {\bibfnamefont {Z.}~\bibnamefont {Xu}}, \bibinfo {author} {\bibfnamefont
  {D.~H.}\ \bibnamefont {Rischke}}, \ and\ \bibinfo {author} {\bibfnamefont
  {C.}~\bibnamefont {Greiner}},\ }\href {\doibase 10.1103/PhysRevD.89.074005}
  {\bibfield  {journal} {\bibinfo  {journal} {Phys. Rev. D}\ }\textbf {\bibinfo
  {volume} {89}},\ \bibinfo {pages} {074005} (\bibinfo {year}
  {2014}{\natexlab{a}})},\ \Eprint {http://arxiv.org/abs/1207.6811}
  {arXiv:1207.6811 [nucl-th]} \BibitemShut {NoStop}%
\bibitem [{\citenamefont {{Wagner}}\ \emph {et~al.}(2022)\citenamefont
  {{Wagner}}, \citenamefont {{Palermo}},\ and\ \citenamefont
  {{Ambru{\c{s}}}}}]{WagnerIReD2022}%
  \BibitemOpen
  \bibfield  {author} {\bibinfo {author} {\bibfnamefont {D.}~\bibnamefont
  {{Wagner}}}, \bibinfo {author} {\bibfnamefont {A.}~\bibnamefont {{Palermo}}},
  \ and\ \bibinfo {author} {\bibfnamefont {V.~E.}\ \bibnamefont
  {{Ambru{\c{s}}}}},\ }\href {\doibase 10.1103/PhysRevD.106.016013} {\bibfield
  {journal} {\bibinfo  {journal} {\prd}\ }\textbf {\bibinfo {volume} {106}},\
  \bibinfo {eid} {016013} (\bibinfo {year} {2022})},\ \Eprint
  {http://arxiv.org/abs/2203.12608} {arXiv:2203.12608 [nucl-th]} \BibitemShut
  {NoStop}%
\bibitem [{\citenamefont {Fotakis}\ \emph {et~al.}(2022)\citenamefont
  {Fotakis}, \citenamefont {Moln\'ar}, \citenamefont {Niemi}, \citenamefont
  {Greiner},\ and\ \citenamefont {Rischke}}]{Fotakis:2022usk}%
  \BibitemOpen
  \bibfield  {author} {\bibinfo {author} {\bibfnamefont {J.~A.}\ \bibnamefont
  {Fotakis}}, \bibinfo {author} {\bibfnamefont {E.}~\bibnamefont {Moln\'ar}},
  \bibinfo {author} {\bibfnamefont {H.}~\bibnamefont {Niemi}}, \bibinfo
  {author} {\bibfnamefont {C.}~\bibnamefont {Greiner}}, \ and\ \bibinfo
  {author} {\bibfnamefont {D.~H.}\ \bibnamefont {Rischke}},\ }\href {\doibase
  10.1103/PhysRevD.106.036009} {\bibfield  {journal} {\bibinfo  {journal}
  {Phys. Rev. D}\ }\textbf {\bibinfo {volume} {106}},\ \bibinfo {pages}
  {036009} (\bibinfo {year} {2022})},\ \Eprint
  {http://arxiv.org/abs/2203.11549} {arXiv:2203.11549 [nucl-th]} \BibitemShut
  {NoStop}%
\bibitem [{\citenamefont {Rocha}\ \emph {et~al.}(2023)\citenamefont {Rocha},
  \citenamefont {de~Brito},\ and\ \citenamefont {Denicol}}]{Rocha:2023hts}%
  \BibitemOpen
  \bibfield  {author} {\bibinfo {author} {\bibfnamefont {G.~S.}\ \bibnamefont
  {Rocha}}, \bibinfo {author} {\bibfnamefont {C.~V.~P.}\ \bibnamefont
  {de~Brito}}, \ and\ \bibinfo {author} {\bibfnamefont {G.~S.}\ \bibnamefont
  {Denicol}},\ }\href {\doibase 10.1103/PhysRevD.108.036017} {\bibfield
  {journal} {\bibinfo  {journal} {Phys. Rev. D}\ }\textbf {\bibinfo {volume}
  {108}},\ \bibinfo {pages} {036017} (\bibinfo {year} {2023})},\ \Eprint
  {http://arxiv.org/abs/2306.07423} {arXiv:2306.07423 [nucl-th]} \BibitemShut
  {NoStop}%
\bibitem [{\citenamefont {Struchtrup}(2004)}]{Struchtrup}%
  \BibitemOpen
  \bibfield  {author} {\bibinfo {author} {\bibfnamefont {H.}~\bibnamefont
  {Struchtrup}},\ }\href {\doibase https://doi.org/10.1063/1.1782751}
  {\bibfield  {journal} {\bibinfo  {journal} {Phys. Fluids}\ }\textbf {\bibinfo
  {volume} {16}},\ \bibinfo {pages} {3921} (\bibinfo {year}
  {2004})}\BibitemShut {NoStop}%
\bibitem [{\citenamefont {Denicol}\ \emph
  {et~al.}(2014{\natexlab{b}})\citenamefont {Denicol}, \citenamefont {Heinz},
  \citenamefont {Martinez}, \citenamefont {Noronha},\ and\ \citenamefont
  {Strickland}}]{Denicol:2014xca}%
  \BibitemOpen
  \bibfield  {author} {\bibinfo {author} {\bibfnamefont {G.~S.}\ \bibnamefont
  {Denicol}}, \bibinfo {author} {\bibfnamefont {U.~W.}\ \bibnamefont {Heinz}},
  \bibinfo {author} {\bibfnamefont {M.}~\bibnamefont {Martinez}}, \bibinfo
  {author} {\bibfnamefont {J.}~\bibnamefont {Noronha}}, \ and\ \bibinfo
  {author} {\bibfnamefont {M.}~\bibnamefont {Strickland}},\ }\href {\doibase
  10.1103/PhysRevLett.113.202301} {\bibfield  {journal} {\bibinfo  {journal}
  {Phys. Rev. Lett.}\ }\textbf {\bibinfo {volume} {113}},\ \bibinfo {pages}
  {202301} (\bibinfo {year} {2014}{\natexlab{b}})},\ \Eprint
  {http://arxiv.org/abs/1408.5646} {arXiv:1408.5646 [hep-ph]} \BibitemShut
  {NoStop}%
\bibitem [{\citenamefont {Strickland}(2014)}]{Strickland:2014pga}%
  \BibitemOpen
  \bibfield  {author} {\bibinfo {author} {\bibfnamefont {M.}~\bibnamefont
  {Strickland}},\ }\href {\doibase 10.5506/APhysPolB.45.2355} {\bibfield
  {journal} {\bibinfo  {journal} {Acta Phys. Polon. B}\ }\textbf {\bibinfo
  {volume} {45}},\ \bibinfo {pages} {2355} (\bibinfo {year} {2014})},\ \Eprint
  {http://arxiv.org/abs/1410.5786} {arXiv:1410.5786 [nucl-th]} \BibitemShut
  {NoStop}%
\bibitem [{\citenamefont {Denicol}\ \emph
  {et~al.}(2014{\natexlab{c}})\citenamefont {Denicol}, \citenamefont {Heinz},
  \citenamefont {Martinez}, \citenamefont {Noronha},\ and\ \citenamefont
  {Strickland}}]{Denicol:2014tha}%
  \BibitemOpen
  \bibfield  {author} {\bibinfo {author} {\bibfnamefont {G.~S.}\ \bibnamefont
  {Denicol}}, \bibinfo {author} {\bibfnamefont {U.~W.}\ \bibnamefont {Heinz}},
  \bibinfo {author} {\bibfnamefont {M.}~\bibnamefont {Martinez}}, \bibinfo
  {author} {\bibfnamefont {J.}~\bibnamefont {Noronha}}, \ and\ \bibinfo
  {author} {\bibfnamefont {M.}~\bibnamefont {Strickland}},\ }\href {\doibase
  10.1103/PhysRevD.90.125026} {\bibfield  {journal} {\bibinfo  {journal} {Phys.
  Rev. D}\ }\textbf {\bibinfo {volume} {90}},\ \bibinfo {pages} {125026}
  (\bibinfo {year} {2014}{\natexlab{c}})},\ \Eprint
  {http://arxiv.org/abs/1408.7048} {arXiv:1408.7048 [hep-ph]} \BibitemShut
  {NoStop}%
\bibitem [{\citenamefont {Alqahtani}\ \emph {et~al.}(2018)\citenamefont
  {Alqahtani}, \citenamefont {Nopoush},\ and\ \citenamefont
  {Strickland}}]{Alqahtani:2017mhy}%
  \BibitemOpen
  \bibfield  {author} {\bibinfo {author} {\bibfnamefont {M.}~\bibnamefont
  {Alqahtani}}, \bibinfo {author} {\bibfnamefont {M.}~\bibnamefont {Nopoush}},
  \ and\ \bibinfo {author} {\bibfnamefont {M.}~\bibnamefont {Strickland}},\
  }\href {\doibase 10.1016/j.ppnp.2018.05.004} {\bibfield  {journal} {\bibinfo
  {journal} {Prog. Part. Nucl. Phys.}\ }\textbf {\bibinfo {volume} {101}},\
  \bibinfo {pages} {204} (\bibinfo {year} {2018})},\ \Eprint
  {http://arxiv.org/abs/1712.03282} {arXiv:1712.03282 [nucl-th]} \BibitemShut
  {NoStop}%
\bibitem [{\citenamefont {Jaiswal}\ \emph {et~al.}(2019)\citenamefont
  {Jaiswal}, \citenamefont {Chattopadhyay}, \citenamefont {Jaiswal},
  \citenamefont {Pal},\ and\ \citenamefont {Heinz}}]{Jaiswal:2019cju}%
  \BibitemOpen
  \bibfield  {author} {\bibinfo {author} {\bibfnamefont {S.}~\bibnamefont
  {Jaiswal}}, \bibinfo {author} {\bibfnamefont {C.}~\bibnamefont
  {Chattopadhyay}}, \bibinfo {author} {\bibfnamefont {A.}~\bibnamefont
  {Jaiswal}}, \bibinfo {author} {\bibfnamefont {S.}~\bibnamefont {Pal}}, \ and\
  \bibinfo {author} {\bibfnamefont {U.}~\bibnamefont {Heinz}},\ }\href
  {\doibase 10.1103/PhysRevC.100.034901} {\bibfield  {journal} {\bibinfo
  {journal} {Phys. Rev. C}\ }\textbf {\bibinfo {volume} {100}},\ \bibinfo
  {pages} {034901} (\bibinfo {year} {2019})},\ \Eprint
  {http://arxiv.org/abs/1907.07965} {arXiv:1907.07965 [nucl-th]} \BibitemShut
  {NoStop}%
\bibitem [{\citenamefont {Dash}\ \emph {et~al.}(2023)\citenamefont {Dash},
  \citenamefont {Shokri}, \citenamefont {Rezzolla},\ and\ \citenamefont
  {Rischke}}]{Dash:2022xkz}%
  \BibitemOpen
  \bibfield  {author} {\bibinfo {author} {\bibfnamefont {A.}~\bibnamefont
  {Dash}}, \bibinfo {author} {\bibfnamefont {M.}~\bibnamefont {Shokri}},
  \bibinfo {author} {\bibfnamefont {L.}~\bibnamefont {Rezzolla}}, \ and\
  \bibinfo {author} {\bibfnamefont {D.~H.}\ \bibnamefont {Rischke}},\ }\href
  {\doibase 10.1103/PhysRevD.107.056003} {\bibfield  {journal} {\bibinfo
  {journal} {Phys. Rev. D}\ }\textbf {\bibinfo {volume} {107}},\ \bibinfo
  {pages} {056003} (\bibinfo {year} {2023})},\ \Eprint
  {http://arxiv.org/abs/2211.09459} {arXiv:2211.09459 [nucl-th]} \BibitemShut
  {NoStop}%
\bibitem [{\citenamefont {{Gavassino}}\ and\ \citenamefont
  {{Noronha}}(2023)}]{GavassinoFarFromBulk2023}%
  \BibitemOpen
  \bibfield  {author} {\bibinfo {author} {\bibfnamefont {L.}~\bibnamefont
  {{Gavassino}}}\ and\ \bibinfo {author} {\bibfnamefont {J.}~\bibnamefont
  {{Noronha}}},\ }\href {\doibase 10.48550/arXiv.2305.04119} {\bibfield
  {journal} {\bibinfo  {journal} {arXiv e-prints}\ ,\ \bibinfo {eid}
  {arXiv:2305.04119}} (\bibinfo {year} {2023})},\ \Eprint
  {http://arxiv.org/abs/2305.04119} {arXiv:2305.04119 [gr-qc]} \BibitemShut
  {NoStop}%
\bibitem [{\citenamefont {{Gavassino}}(2023)}]{GavassinoBurgers2023}%
  \BibitemOpen
  \bibfield  {author} {\bibinfo {author} {\bibfnamefont {L.}~\bibnamefont
  {{Gavassino}}},\ }\href {\doibase 10.48550/arXiv.2304.05455} {\bibfield
  {journal} {\bibinfo  {journal} {arXiv e-prints}\ ,\ \bibinfo {eid}
  {arXiv:2304.05455}} (\bibinfo {year} {2023})},\ \Eprint
  {http://arxiv.org/abs/2304.05455} {arXiv:2304.05455 [nucl-th]} \BibitemShut
  {NoStop}%
\bibitem [{\citenamefont {{Peliti}}(2011)}]{peliti_book}%
  \BibitemOpen
  \bibfield  {author} {\bibinfo {author} {\bibfnamefont {L.}~\bibnamefont
  {{Peliti}}},\ }\href@noop {} {\emph {\bibinfo {title} {{Statistical Mechanics
  in a Nutshell}}}},\ In a nutshell\ (\bibinfo  {publisher} {Princeton
  University Press},\ \bibinfo {year} {2011})\BibitemShut {NoStop}%
\bibitem [{\citenamefont {Heller}\ \emph {et~al.}(2014)\citenamefont {Heller},
  \citenamefont {Janik}, \citenamefont {Spali\ifmmode~\acute{n}\else
  \'{n}\fi{}ski},\ and\ \citenamefont {Witaszczyk}}]{Heller2014}%
  \BibitemOpen
  \bibfield  {author} {\bibinfo {author} {\bibfnamefont {M.~P.}\ \bibnamefont
  {Heller}}, \bibinfo {author} {\bibfnamefont {R.~A.}\ \bibnamefont {Janik}},
  \bibinfo {author} {\bibfnamefont {M.}~\bibnamefont
  {Spali\ifmmode~\acute{n}\else \'{n}\fi{}ski}}, \ and\ \bibinfo {author}
  {\bibfnamefont {P.}~\bibnamefont {Witaszczyk}},\ }\href {\doibase
  10.1103/PhysRevLett.113.261601} {\bibfield  {journal} {\bibinfo  {journal}
  {Phys. Rev. Lett.}\ }\textbf {\bibinfo {volume} {113}},\ \bibinfo {pages}
  {261601} (\bibinfo {year} {2014})}\BibitemShut {NoStop}%
\bibitem [{\citenamefont {Bemfica}\ \emph {et~al.}(2018)\citenamefont
  {Bemfica}, \citenamefont {Disconzi},\ and\ \citenamefont
  {Noronha}}]{Bemfica2017TheFirst}%
  \BibitemOpen
  \bibfield  {author} {\bibinfo {author} {\bibfnamefont {F.~S.}\ \bibnamefont
  {Bemfica}}, \bibinfo {author} {\bibfnamefont {M.~M.}\ \bibnamefont
  {Disconzi}}, \ and\ \bibinfo {author} {\bibfnamefont {J.}~\bibnamefont
  {Noronha}},\ }\href {\doibase 10.1103/PhysRevD.98.104064} {\bibfield
  {journal} {\bibinfo  {journal} {Phys. Rev. D}\ }\textbf {\bibinfo {volume}
  {98}},\ \bibinfo {pages} {104064} (\bibinfo {year} {2018})},\ \Eprint
  {http://arxiv.org/abs/1708.06255} {arXiv:1708.06255 [gr-qc]} \BibitemShut
  {NoStop}%
\bibitem [{\citenamefont {{Kovtun}}(2019)}]{Kovtun2019}%
  \BibitemOpen
  \bibfield  {author} {\bibinfo {author} {\bibfnamefont {P.}~\bibnamefont
  {{Kovtun}}},\ }\href {\doibase 10.1007/JHEP10(2019)034} {\bibfield  {journal}
  {\bibinfo  {journal} {Journal of High Energy Physics}\ }\textbf {\bibinfo
  {volume} {2019}},\ \bibinfo {eid} {34} (\bibinfo {year} {2019})},\ \Eprint
  {http://arxiv.org/abs/1907.08191} {arXiv:1907.08191 [hep-th]} \BibitemShut
  {NoStop}%
\bibitem [{\citenamefont {Bemfica}\ \emph {et~al.}(2022)\citenamefont
  {Bemfica}, \citenamefont {Disconzi},\ and\ \citenamefont
  {Noronha}}]{BemficaDNDefinitivo2020}%
  \BibitemOpen
  \bibfield  {author} {\bibinfo {author} {\bibfnamefont {F.~S.}\ \bibnamefont
  {Bemfica}}, \bibinfo {author} {\bibfnamefont {M.~M.}\ \bibnamefont
  {Disconzi}}, \ and\ \bibinfo {author} {\bibfnamefont {J.}~\bibnamefont
  {Noronha}},\ }\href {\doibase 10.1103/PhysRevX.12.021044} {\bibfield
  {journal} {\bibinfo  {journal} {Phys. Rev. X}\ }\textbf {\bibinfo {volume}
  {12}},\ \bibinfo {pages} {021044} (\bibinfo {year} {2022})}\BibitemShut
  {NoStop}%
\bibitem [{\citenamefont {{Camelio}}\ \emph {et~al.}(2022)\citenamefont
  {{Camelio}}, \citenamefont {{Gavassino}}, \citenamefont {{Antonelli}},
  \citenamefont {{Bernuzzi}},\ and\ \citenamefont {{Haskell}}}]{Camelio2022}%
  \BibitemOpen
  \bibfield  {author} {\bibinfo {author} {\bibfnamefont {G.}~\bibnamefont
  {{Camelio}}}, \bibinfo {author} {\bibfnamefont {L.}~\bibnamefont
  {{Gavassino}}}, \bibinfo {author} {\bibfnamefont {M.}~\bibnamefont
  {{Antonelli}}}, \bibinfo {author} {\bibfnamefont {S.}~\bibnamefont
  {{Bernuzzi}}}, \ and\ \bibinfo {author} {\bibfnamefont {B.}~\bibnamefont
  {{Haskell}}},\ }\href@noop {} {\bibfield  {journal} {\bibinfo  {journal}
  {arXiv e-prints}\ ,\ \bibinfo {eid} {arXiv:2204.11809}} (\bibinfo {year}
  {2022})},\ \Eprint {http://arxiv.org/abs/2204.11809} {arXiv:2204.11809
  [gr-qc]} \BibitemShut {NoStop}%
\bibitem [{\citenamefont {de~Groot}\ \emph {et~al.}(1980)\citenamefont
  {de~Groot}, \citenamefont {van Leeuwen},\ and\ \citenamefont {van
  Weert}}]{DeGroot:1980dk}%
  \BibitemOpen
  \bibfield  {author} {\bibinfo {author} {\bibfnamefont {S.~R.}\ \bibnamefont
  {de~Groot}}, \bibinfo {author} {\bibfnamefont {W.~A.}\ \bibnamefont {van
  Leeuwen}}, \ and\ \bibinfo {author} {\bibfnamefont {C.~G.}\ \bibnamefont {van
  Weert}},\ }\href@noop {} {\emph {\bibinfo {title} {Relativistic Kinetic
  Theory. Principles and Applications}}}\ (\bibinfo  {publisher}
  {North-Holland},\ \bibinfo {year} {1980})\BibitemShut {NoStop}%
\bibitem [{\citenamefont {Cercignani}\ and\ \citenamefont
  {Kremer}(2002)}]{Cercignani}%
  \BibitemOpen
  \bibfield  {author} {\bibinfo {author} {\bibfnamefont {C.}~\bibnamefont
  {Cercignani}}\ and\ \bibinfo {author} {\bibfnamefont {G.~M.}\ \bibnamefont
  {Kremer}},\ }\href@noop {} {\emph {\bibinfo {title} {The Relativistic
  Boltzmann Equation: Theory and Applications}}}\ (\bibinfo  {publisher}
  {Birkh\"{a}user},\ \bibinfo {year} {2002})\BibitemShut {NoStop}%
\bibitem [{\citenamefont {Wagner}\ \emph {et~al.}(2023)\citenamefont {Wagner},
  \citenamefont {Ambrus},\ and\ \citenamefont {Moln\'ar}}]{Wagner:2023joq}%
  \BibitemOpen
  \bibfield  {author} {\bibinfo {author} {\bibfnamefont {D.}~\bibnamefont
  {Wagner}}, \bibinfo {author} {\bibfnamefont {V.~E.}\ \bibnamefont {Ambrus}},
  \ and\ \bibinfo {author} {\bibfnamefont {E.}~\bibnamefont {Moln\'ar}},\
  }\href@noop {} {\  (\bibinfo {year} {2023})},\ \Eprint
  {http://arxiv.org/abs/2309.09335} {arXiv:2309.09335 [physics.flu-dyn]}
  \BibitemShut {NoStop}%
\bibitem [{\citenamefont {Ambrus}\ \emph {et~al.}(2022)\citenamefont {Ambrus},
  \citenamefont {Moln\'ar},\ and\ \citenamefont {Rischke}}]{Ambrus:2022vif}%
  \BibitemOpen
  \bibfield  {author} {\bibinfo {author} {\bibfnamefont {V.~E.}\ \bibnamefont
  {Ambrus}}, \bibinfo {author} {\bibfnamefont {E.}~\bibnamefont {Moln\'ar}}, \
  and\ \bibinfo {author} {\bibfnamefont {D.~H.}\ \bibnamefont {Rischke}},\
  }\href {\doibase 10.1103/PhysRevD.106.076005} {\bibfield  {journal} {\bibinfo
   {journal} {Phys. Rev. D}\ }\textbf {\bibinfo {volume} {106}},\ \bibinfo
  {pages} {076005} (\bibinfo {year} {2022})},\ \Eprint
  {http://arxiv.org/abs/2207.05670} {arXiv:2207.05670 [nucl-th]} \BibitemShut
  {NoStop}%
\bibitem [{\citenamefont {{Kovtun}}\ \emph {et~al.}(2011)\citenamefont
  {{Kovtun}}, \citenamefont {{Moore}},\ and\ \citenamefont
  {{Romatschke}}}]{KovtunStickiness2011}%
  \BibitemOpen
  \bibfield  {author} {\bibinfo {author} {\bibfnamefont {P.}~\bibnamefont
  {{Kovtun}}}, \bibinfo {author} {\bibfnamefont {G.~D.}\ \bibnamefont
  {{Moore}}}, \ and\ \bibinfo {author} {\bibfnamefont {P.}~\bibnamefont
  {{Romatschke}}},\ }\href {\doibase 10.1103/PhysRevD.84.025006} {\bibfield
  {journal} {\bibinfo  {journal} {\prd}\ }\textbf {\bibinfo {volume} {84}},\
  \bibinfo {eid} {025006} (\bibinfo {year} {2011})},\ \Eprint
  {http://arxiv.org/abs/1104.1586} {arXiv:1104.1586 [hep-ph]} \BibitemShut
  {NoStop}%
\end{thebibliography}%

\label{lastpage}

\end{document}